\documentclass[12pt]{article}
\usepackage{latexsym,amsmath,amsthm,amssymb,amscd, bbm, graphicx}
\usepackage[mathscr]{euscript}

\usepackage{float}
\usepackage{caption}
\usepackage[round]{natbib}
\usepackage{apalike}
\usepackage{authblk}

\usepackage[pdftex]{hyperref}
\usepackage{color}
\usepackage{fullpage}
\usepackage{setspace}
\usepackage{url}

\date{}\title{Ireland Topsoil Contamination Analysis: A Clustering Approach}
\usepackage{authblk}
\author{\normalsize{Mimi Zhang}}
\affil{\small{School of Computer Science and Statistics, Trinity College Dublin, Ireland}}

\begin{document}
\maketitle

\section{Background}
In July 2023, the European Commission proposed the Soil Monitoring Law to standardize systematic soil monitoring across the EU and mandate the creation of national contaminated site registries \citep{European_Commission_2023}. However, Ireland faces significant challenges in meeting the five-year implementation deadline, primarily due to the lack of a centralized national registry—a core requirement of the legislation. Compounding this issue, fewer than one-third of Irish soil studies align with research priorities outlined in the EU Soil Strategy for 2030 and national policies \citep{EPA_2022}. Benchmarking Irish research against European standards reveals a critical need for more standardized approaches to soil contamination assessment to comply with the new regulatory framework.

This study analyzes shallow topsoil geochemical data from the Tellus Programme \citep{GSI_Tellus}, a national mapping initiative led by the Geological Survey Ireland (GSI). The Tellus dataset provides a valuable yet underutilized resource for addressing Ireland’s soil contamination knowledge gaps. While previous applications have been limited to single-element analyses or localized regional studies, this research employs multivariate and machine learning techniques to uncover complex interactions among potentially toxic elements (PTEs). The data and reproducible code used to generate the clusters are openly available in the GitHub repository: \url{https://github.com/tobinjo96/CPFcluster/tree/master/Spatial-CPF}. Due to size constraints, the datasets and scripts for visualizing soil samples on bedrock and land cover maps could not be hosted directly but are available upon request.

\section{Data Preparation}
\begin{figure}[htbp]
	\centering
	\includegraphics[width=\textwidth]{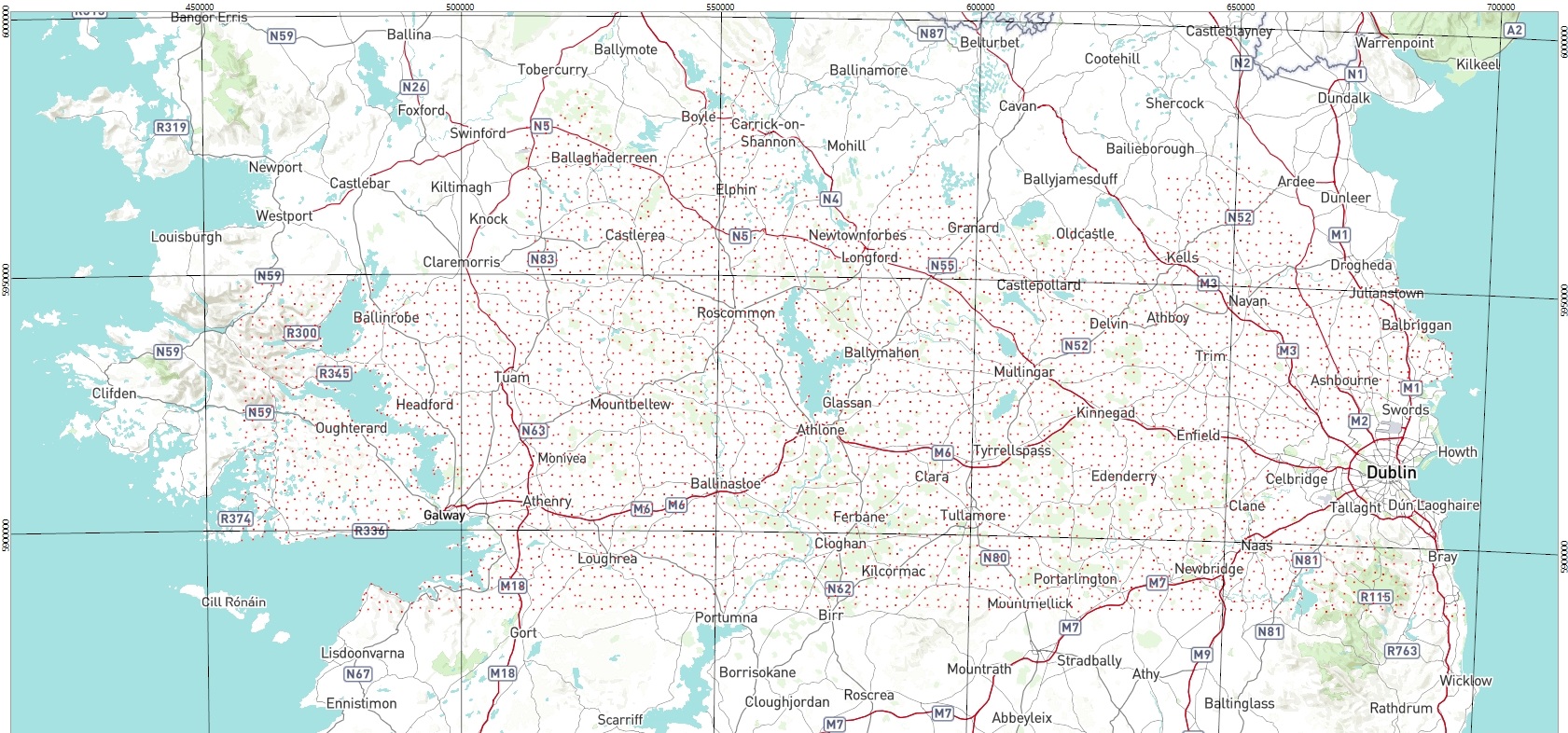}
	\caption{The sampling sites (red dots) of the ``G5'' dataset.}
\label{sampling_sites_dot}
\end{figure}

The dataset, referred to as ``G5'', was collected between 2017 and 2019 and is publicly available \href{https://www.gsi.ie/en-ie/data-and-maps/Pages/Geochemistry.aspx}{here}. It covers 17,983 km² (24.3\% of the country), encompassing the West, Midlands, and East of Ireland; see Figure \ref{sampling_sites_dot}.
These regions were selected for their diverse geological formations and land uses, which significantly influence the distribution and mobility of PTEs. Notably, prior soil studies have predominantly focused on Dublin and Northern Ireland \citep{EPA_2022}, leaving other regions understudied. This study addresses this gap by leveraging the GSI Tellus datasets. The dataset comprises 4,278 topsoil samples collected at a density of one site per 4 km² and depths of 5–20 cm. The  samples  underwent  multi-element  partial  extract  analysis  using inductively coupled plasma mass spectrometry (ICP-MS) followed by aqua regia digestion in accredited facilities. During ICP-MS analysis, certain elements (Ta, Au, Pd, Pt, and Re) were excluded due to $>5\%$ of values falling below the detection limit. For this study, 15 key PTEs were selected (measured in mg/kg): As, Ba, Bi, Co, Cr, Cu, Mn, Mo, Ni, Pb, Sb, Sn, U, V, and Zn. These elements were chosen based on three criteria: (1) their high contamination risk in soils \citep{REIMANN2018302}, (2) documented harmful effects on human health and the environment, and (3) historical prevalence in Irish soils. This targeted selection ensures a focused analysis of the most significant threats to soil quality.

Coordinates were transformed from the Irish Transverse Mercator (ITM, EPSG:2157) system to the global WGS84 system (EPSG:4326) to ensure compatibility with web mapping libraries (e.g., Folium) and GIS tools. The original and processed datasets are both available at \url{https://github.com/tobinjo96/CPFcluster/tree/master/Spatial-CPF/Data}. A bedrock geological map (Shapefile format) at a scale of 1:1,000,000 was obtained from the GSI website. The land cover map was generated using data from the European Union's Copernicus Land Monitoring Service, with reference to the year 2018.

\section{Methodology}
In this study, we apply the CPF clustering method \citep{10296014} to classify soil samples into distinct groups and analyze topsoil contamination patterns across Ireland. The method is implemented in the publicly available Python package CPFcluster (\url{https://github.com/tobinjo96/CPFcluster}). Given the inclusion of geographic coordinates in our dataset, we utilize the functions in the Spatial-CPF folder, which incorporates spatial constraints into the clustering process.

We begin by running the \textit{create\_neighbor\_adjacency\_matrix} function to construct the adjacency matrix (namely, the mutual k-nearest neighbor graph) from the geographic coordinates. Due to its large size, the precomputed adjacency matrix file \textit{geo\_neighbor\_adjacency\_matrix.npy} is not included in the Data folder. Users must run the provided code to generate this matrix.  The returned adjacency matrix and the geochemical contamination data are then fed into the \textit{cpf.fit} function.\footnote{The \textit{build\_CCgraph} in the \textit{core\_Geo} file will construct another adjacency matrix based solely on the geochemical contamination data, applying the same k-nearest neighbor threshold as the spatial proximity matrix. The final adjacency matrix is computed as the element-wise product of these two matrices (geographic and geochemical). Thus, two samples are considered neighbors only if they are both geochemically similar and geographically proximate. The final adjacency matrix is used to identify the connected components.} As suggested by \cite{10296014}, we investigate the hyperparameter space near \(min\_samples=\sqrt{n}\approx70\). Through extensive testing of various parameter configurations, we identified the optimal hyperparameter set: \textit{min\_samples=75, rho=0.01, alpha=0.015, merge\_threshold=7.5, density\_ratio\_threshold=0.7}, at which the Calinski-Harabasz score is 76.5202. Users can test alternative hyperparameters by running the \textit{Code\_Example\_Geo.py} script. The clustering results are visualized in Figure \ref{clustering_visualization} using dimensionality reduction techniques:
\begin{figure}[htbp]
	\centering
	\includegraphics[width=0.475\textwidth]{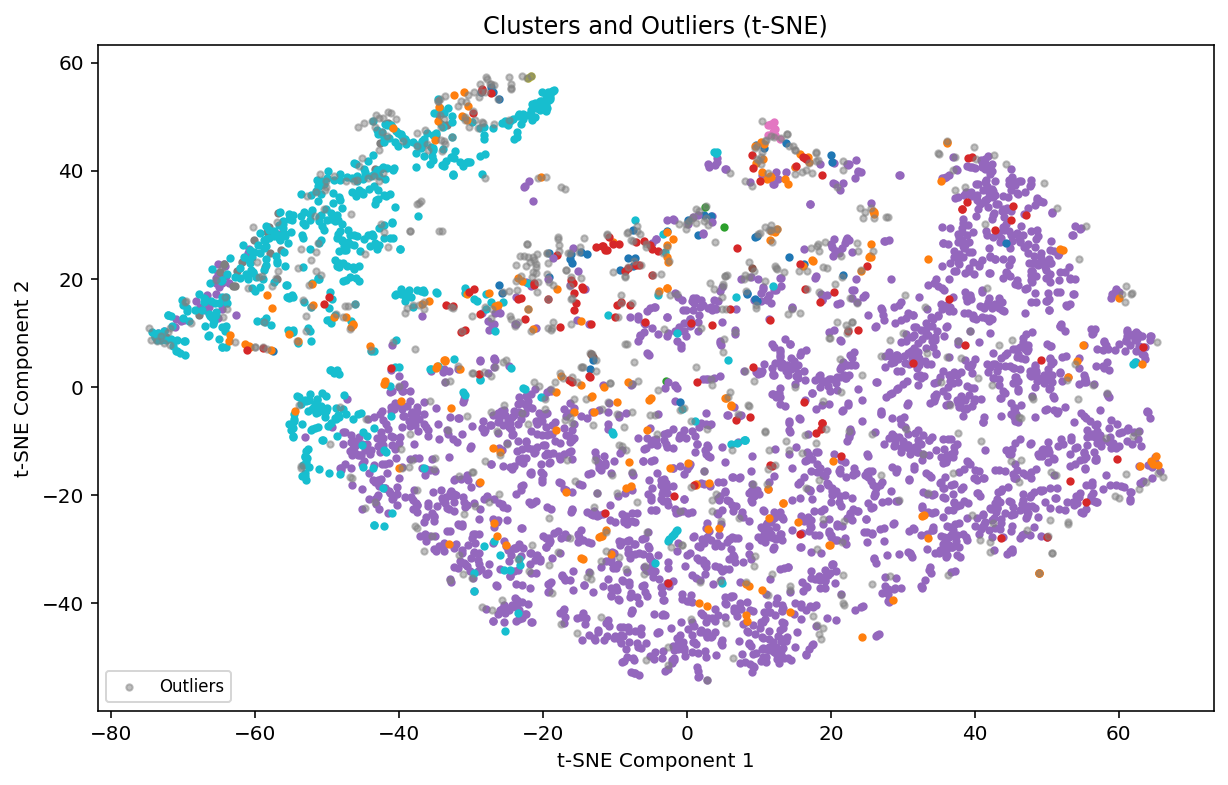}
    \hspace{\fill}
    \includegraphics[width=0.475\textwidth]{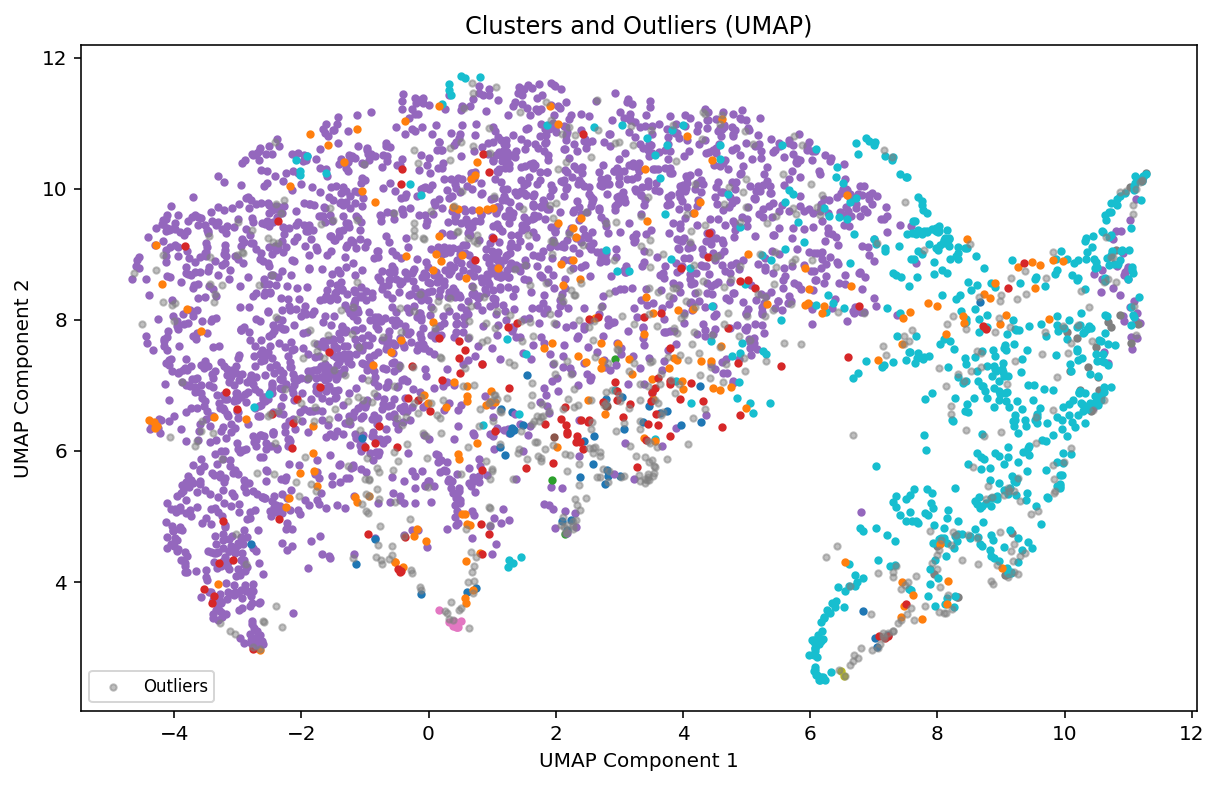}
	\caption{Visualization of the clustering results through t-SNE projection (left) and UMAP projection (right). Different colors represent different clusters.}
\label{clustering_visualization}
\end{figure}

\section{Results}
The CPF algorithm produced eight clusters and a set of 682 outliers, with cluster sizes detailed in Table \ref{tab:results}. Cluster 1 (represented by purple dots in Figure \ref{clustering_visualization}) constitutes the dominant group and contains healthy soil samples, while the smaller clusters (7 and 8) and the outlier set are of particular interest due to their potential to reveal anomalous soil conditions, as analyzed in the following section.
\begin{table}[!ht]
\centering
\caption{Clustering results on the G5 contamination data}
\label{tab:results}
\begin{tabular}{l|cccc}
\hline
     & Cluster 1& Cluster 2& Cluster 3& Cluster 4\\
Size & 2623& 604 & 173& 114\\
     \hline
     & Cluster 5& Cluster 6& Cluster 7& Cluster 8\\
Size & 40& 25& 9& 8\\
\hline
\end{tabular}
\end{table}

\begin{figure}[!h]
	\centering
	\includegraphics[width=0.7\textwidth]{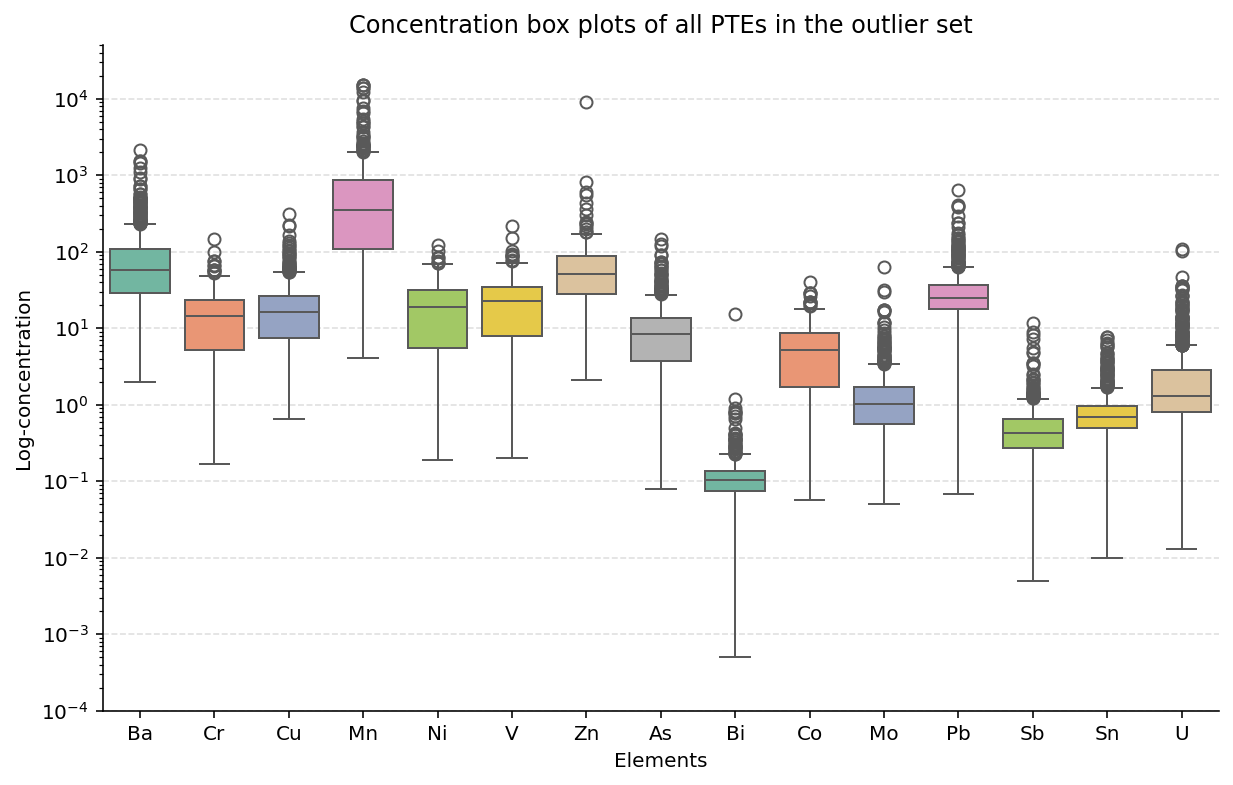}
    \hspace{\fill}
    \includegraphics[width=0.7\textwidth]{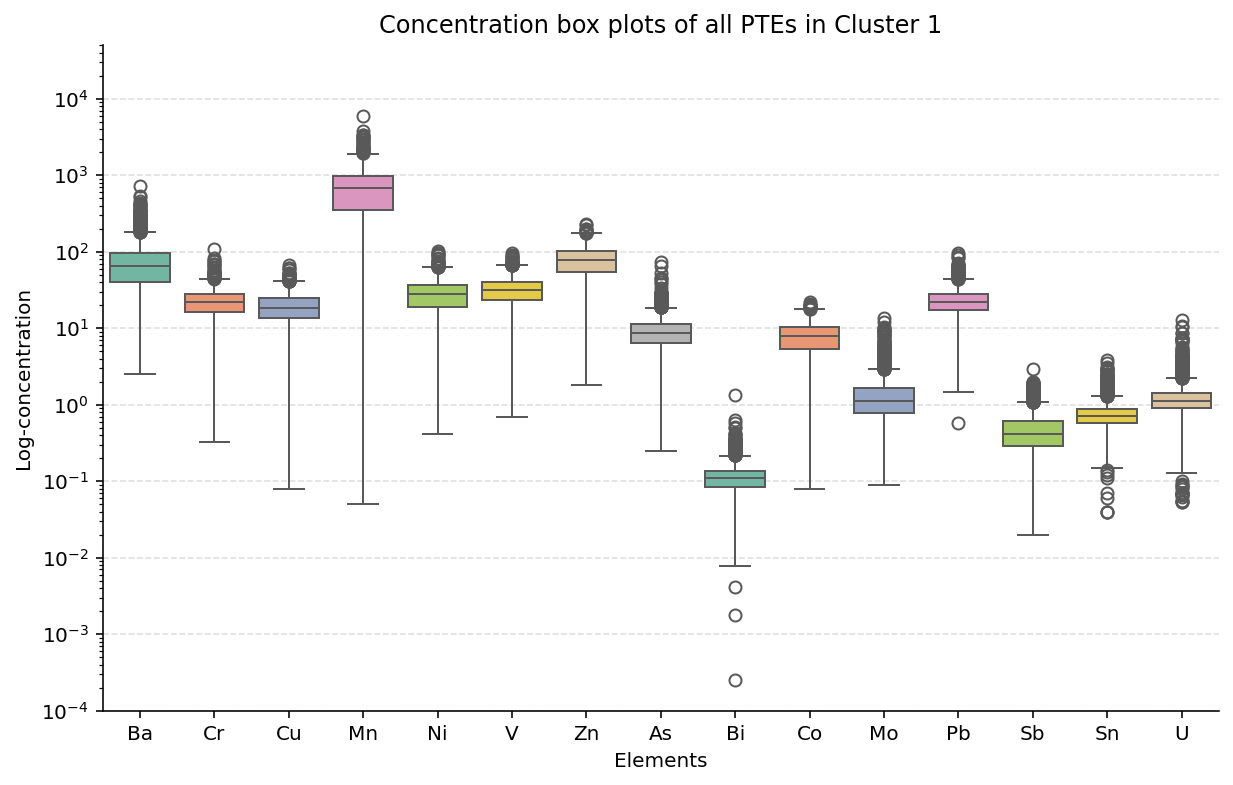}
        \hspace{\fill}
    \includegraphics[width=0.7\textwidth]{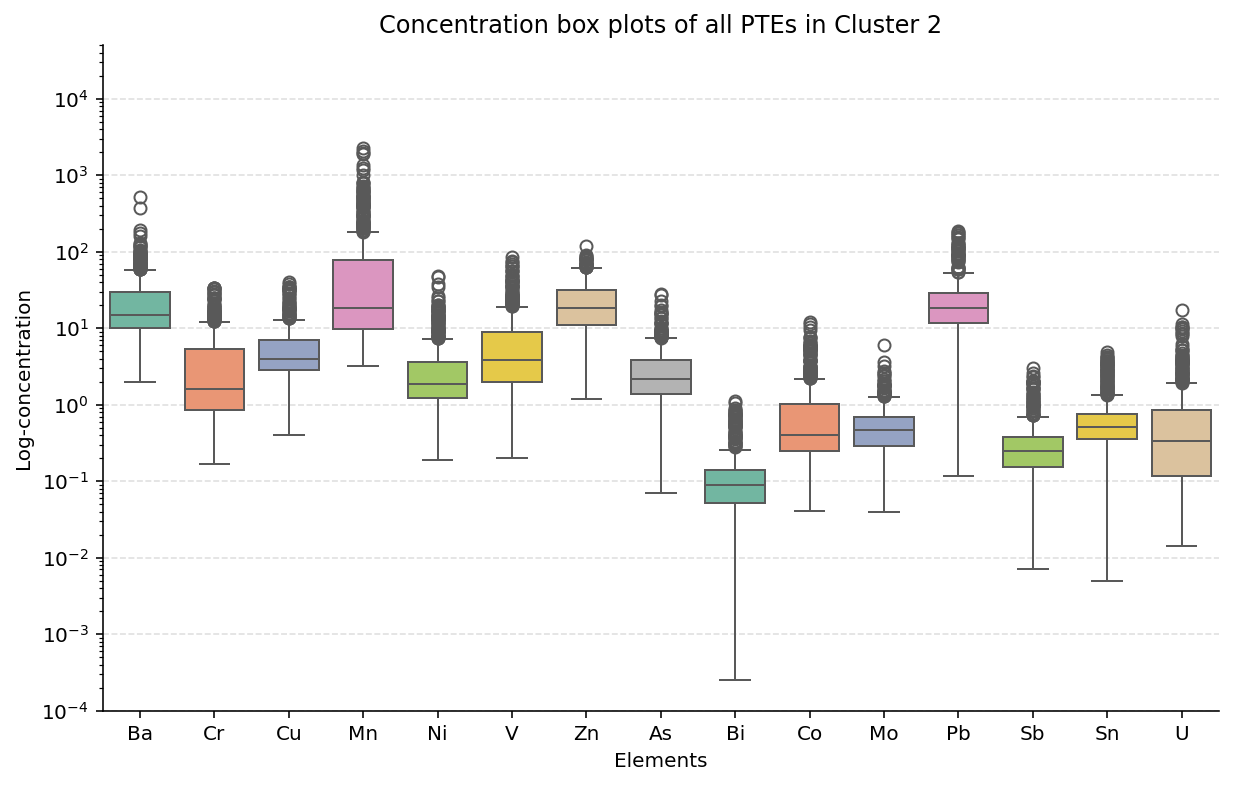}
	\caption{Comparison of PTE concentrations between outlier set (top) and Cluster 1 (bottom). Note the log$_{10}$-transformed y-axis scales, necessary to visualize extreme Mn concentrations in the outliers.}
\label{compare_C1_ourlier}
\end{figure}

Figure \ref{compare_C1_ourlier} compares the PTE concentrations among the three largest subsets: Cluster 1, Cluster 2, and the outlier set, with log-transformed y-axes to accommodate the extreme Mn and Zn values in the outlier set.
It can be observed from Figure \ref{compare_C1_ourlier} that Cluster 1 exhibits the smallest interquartile ranges (IQRs), indicating the lowest variability in PTE concentrations among the three subsets. Cluster 2 shows the lowest median values and lowest upper whiskers, suggesting consistently lower concentrations. In contrast, the outlier set displays both the highest medians and largest IQRs, reflecting elevated concentrations with high variability, along with numerous extreme upper outliers.

\begin{figure}[htbp]
	\centering
    \includegraphics[width=0.475\textwidth]{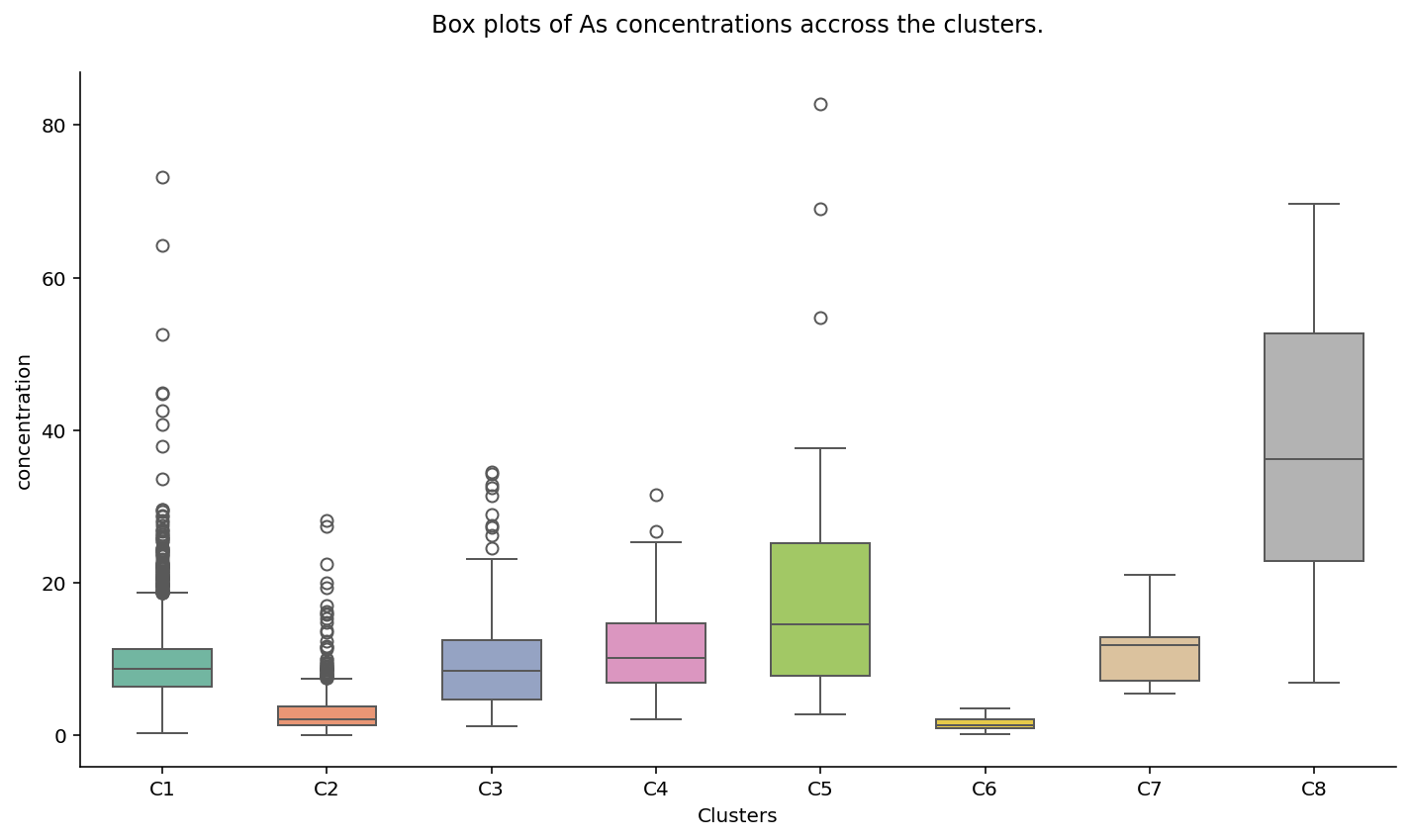}
    \includegraphics[width=0.475\textwidth]{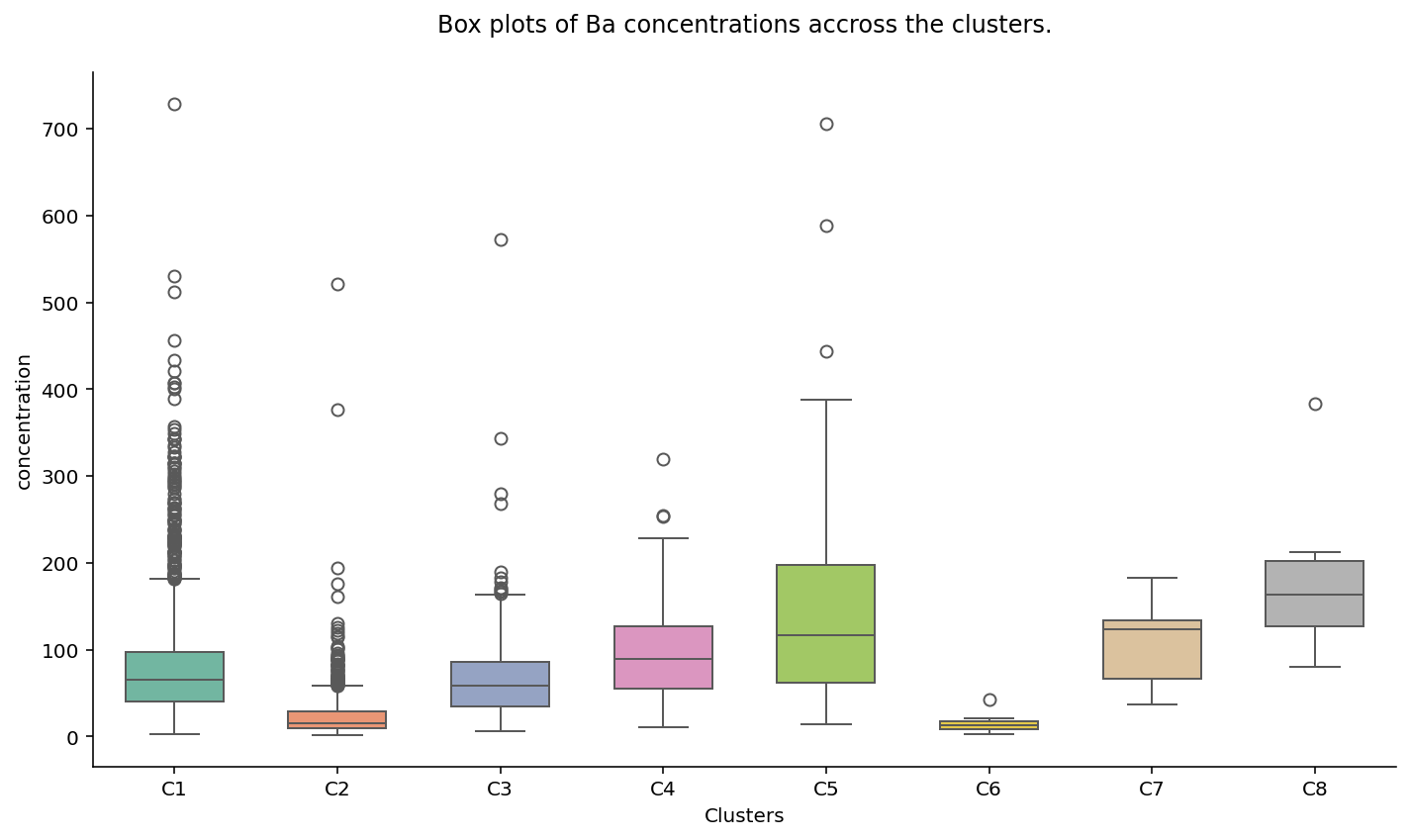}
    \includegraphics[width=0.475\textwidth]{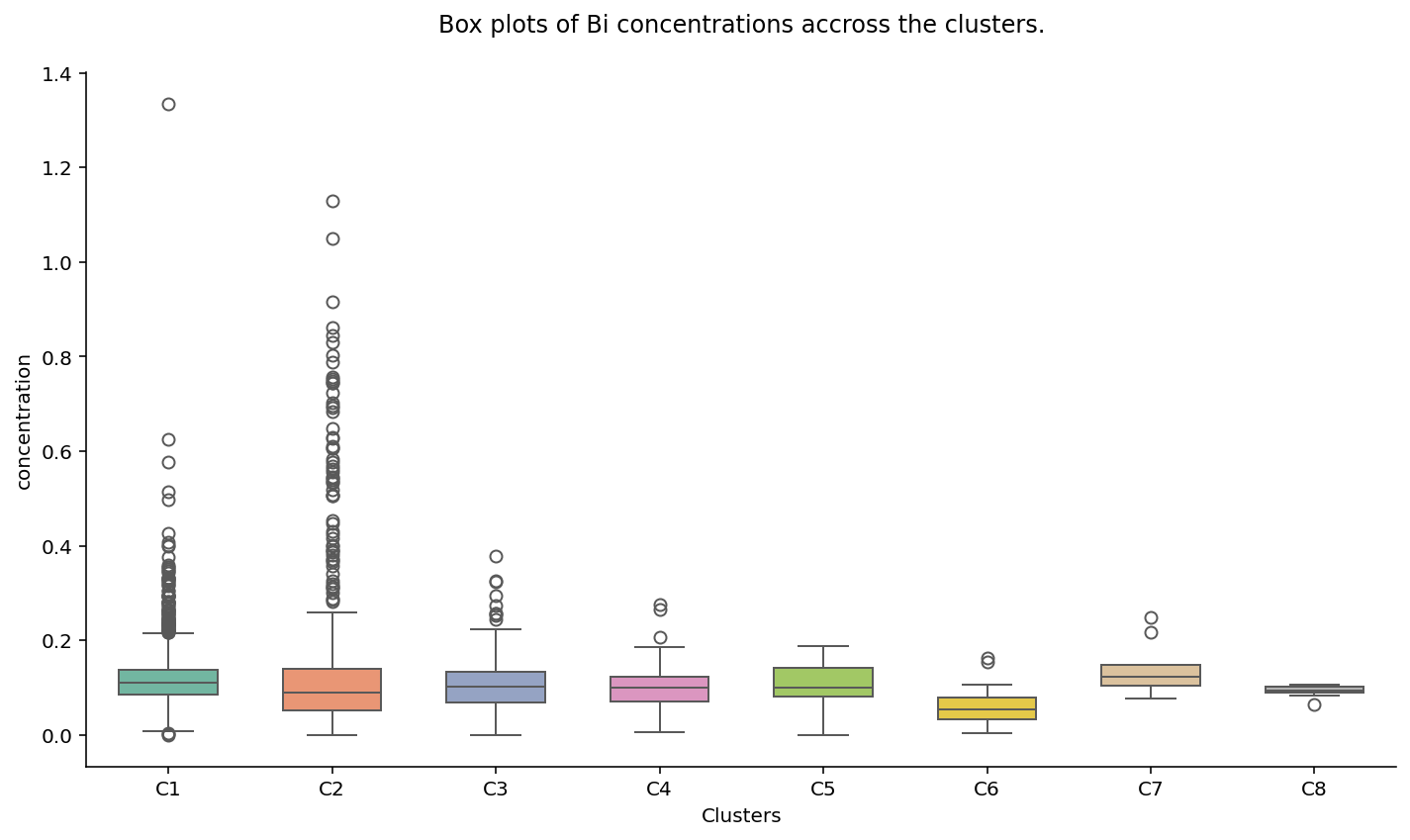}
    \includegraphics[width=0.475\textwidth]{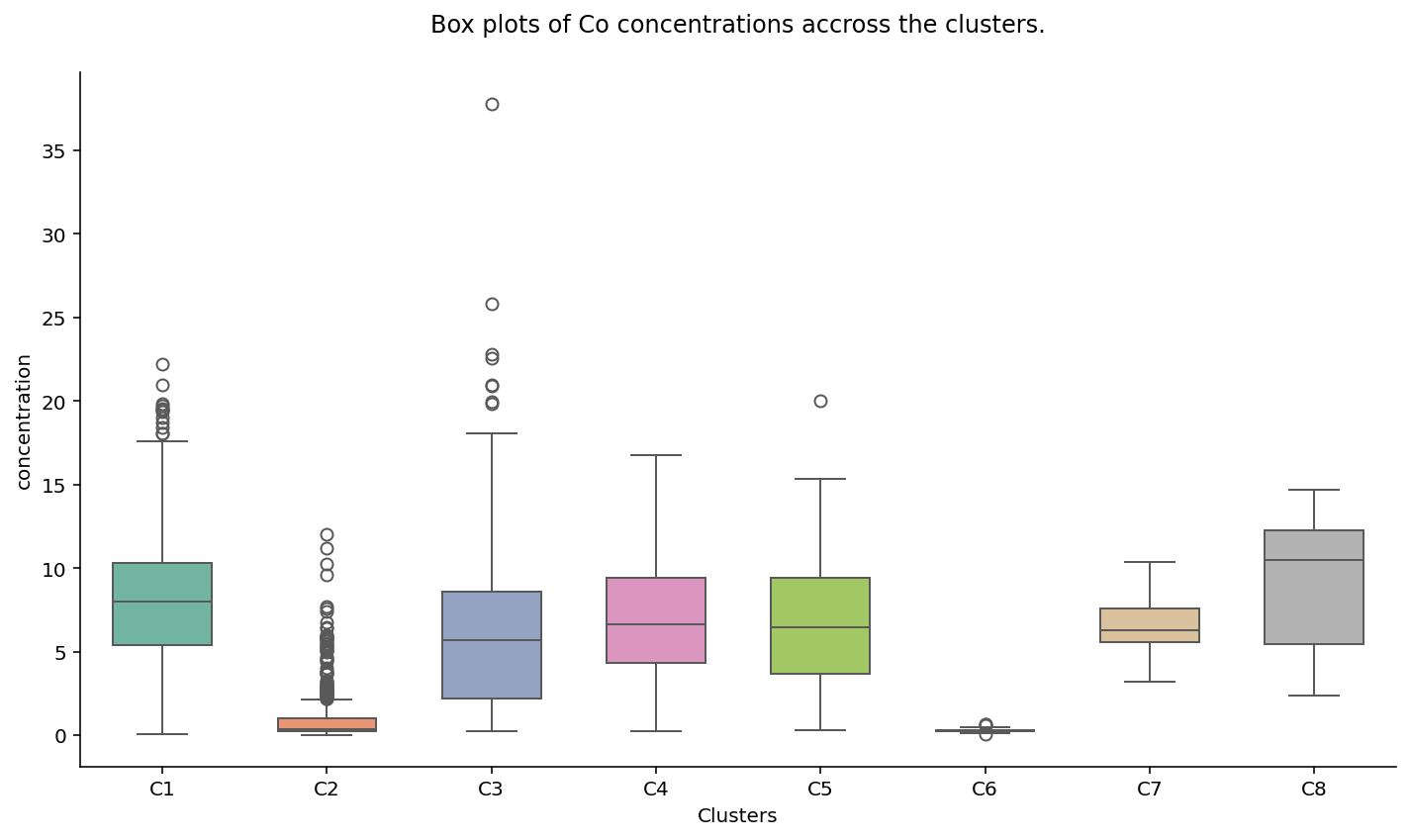}
    \includegraphics[width=0.475\textwidth]{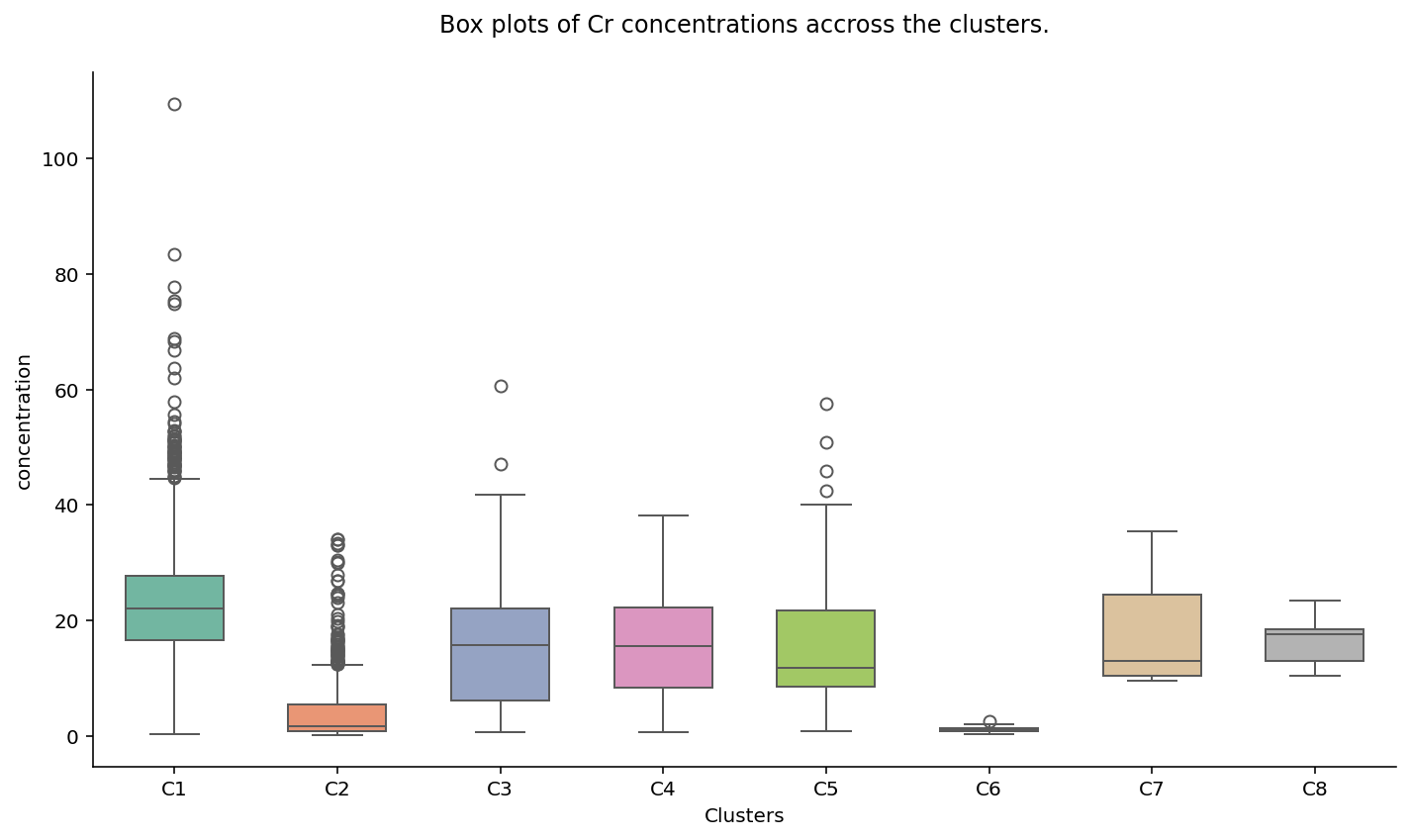}
    \includegraphics[width=0.475\textwidth]{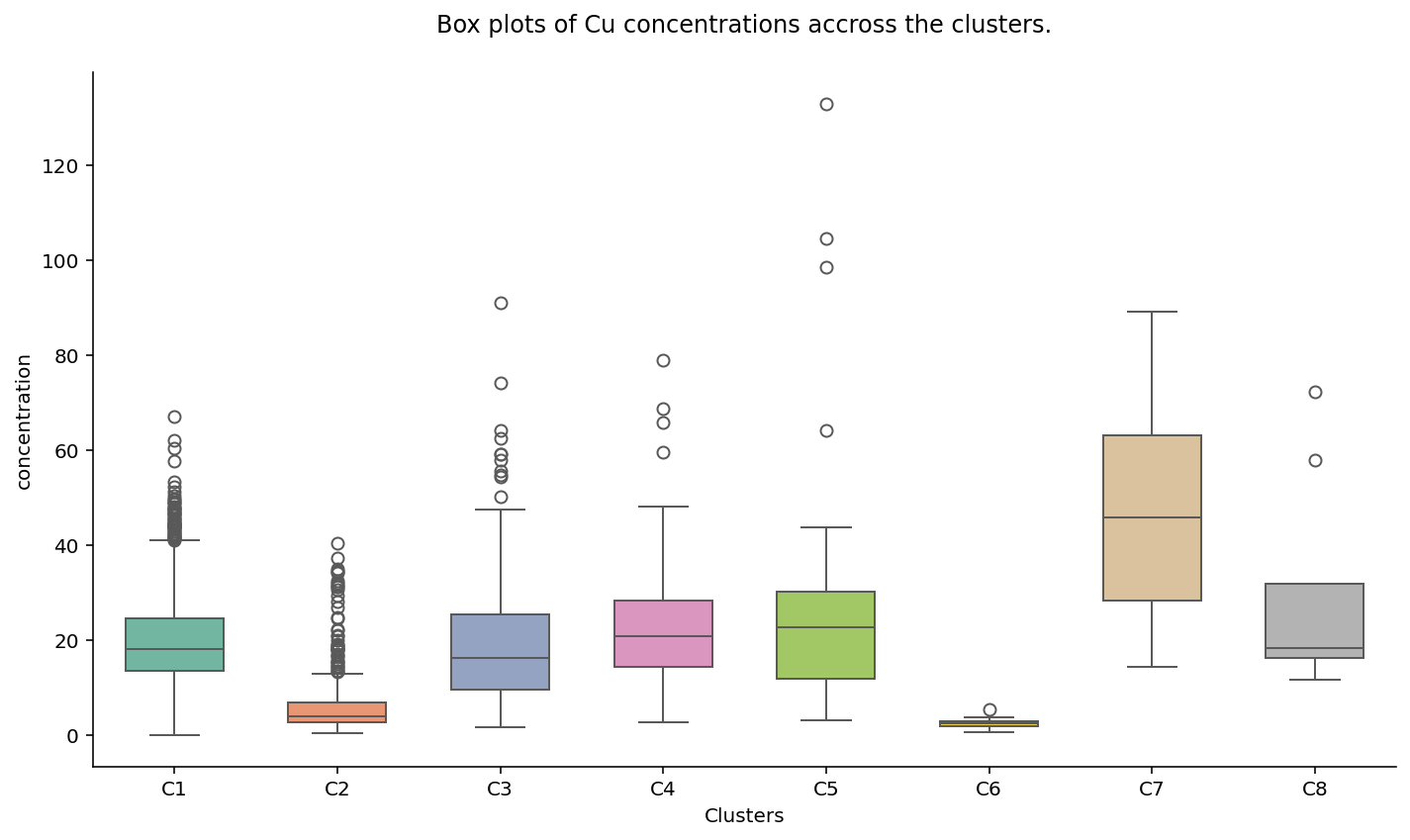}
    \includegraphics[width=0.475\textwidth]{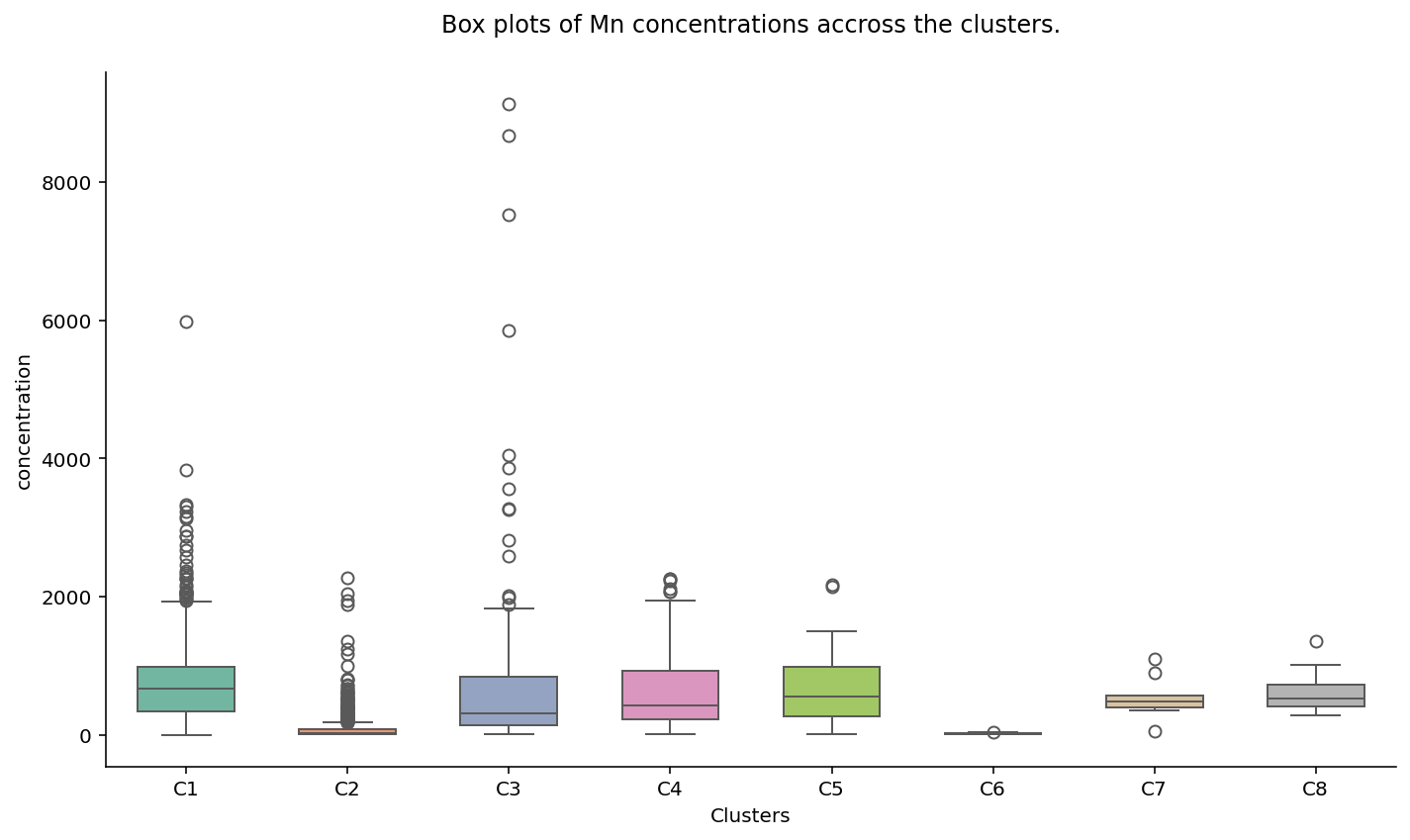}
    \includegraphics[width=0.475\textwidth]{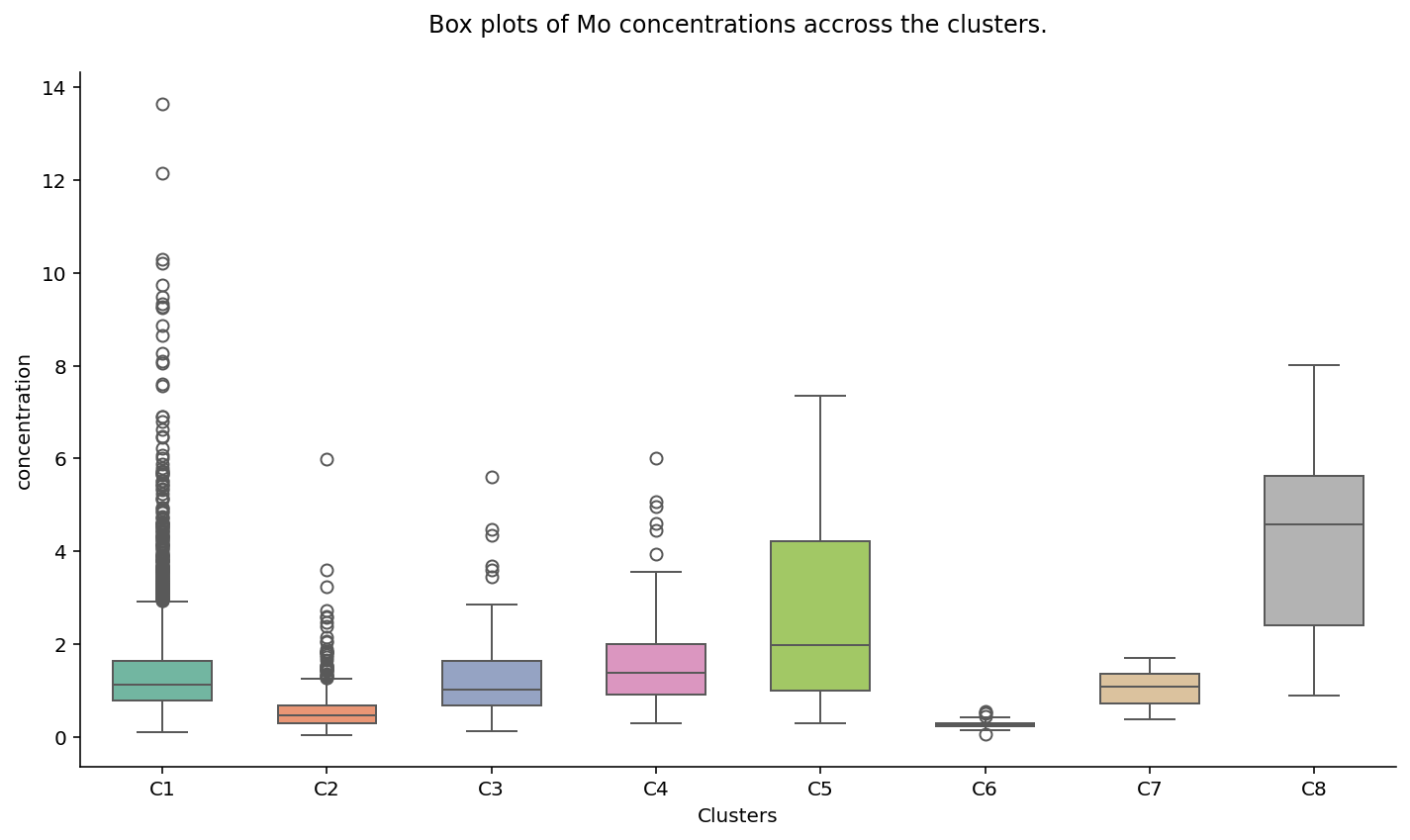}
	\caption{Box plots showing concentration distributions of the 15 PTEs across the eight identified soil clusters.}
\label{boxplot_all1}
\end{figure}

\begin{figure}[htbp]
	\centering
    \includegraphics[width=0.475\textwidth]{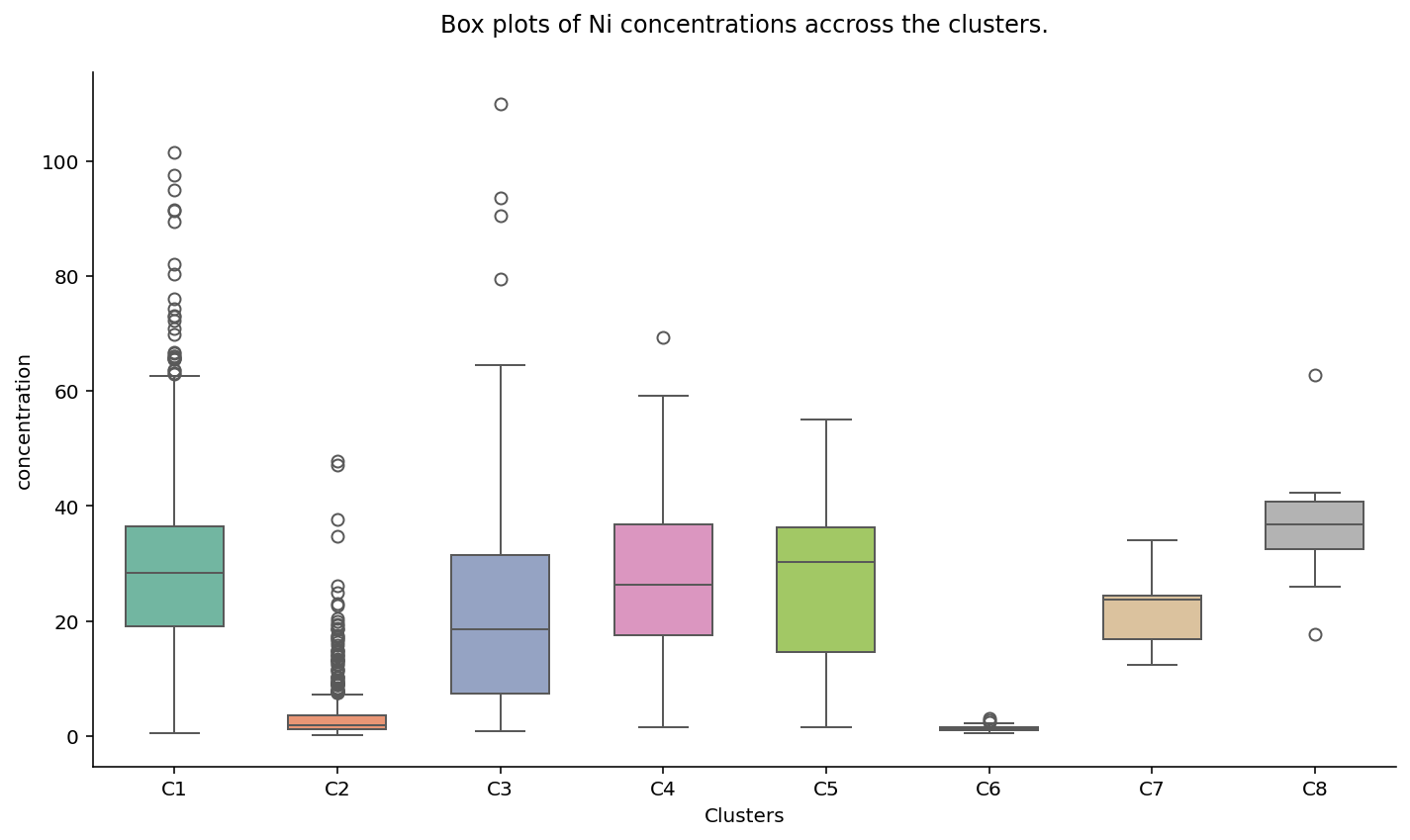}
    \includegraphics[width=0.475\textwidth]{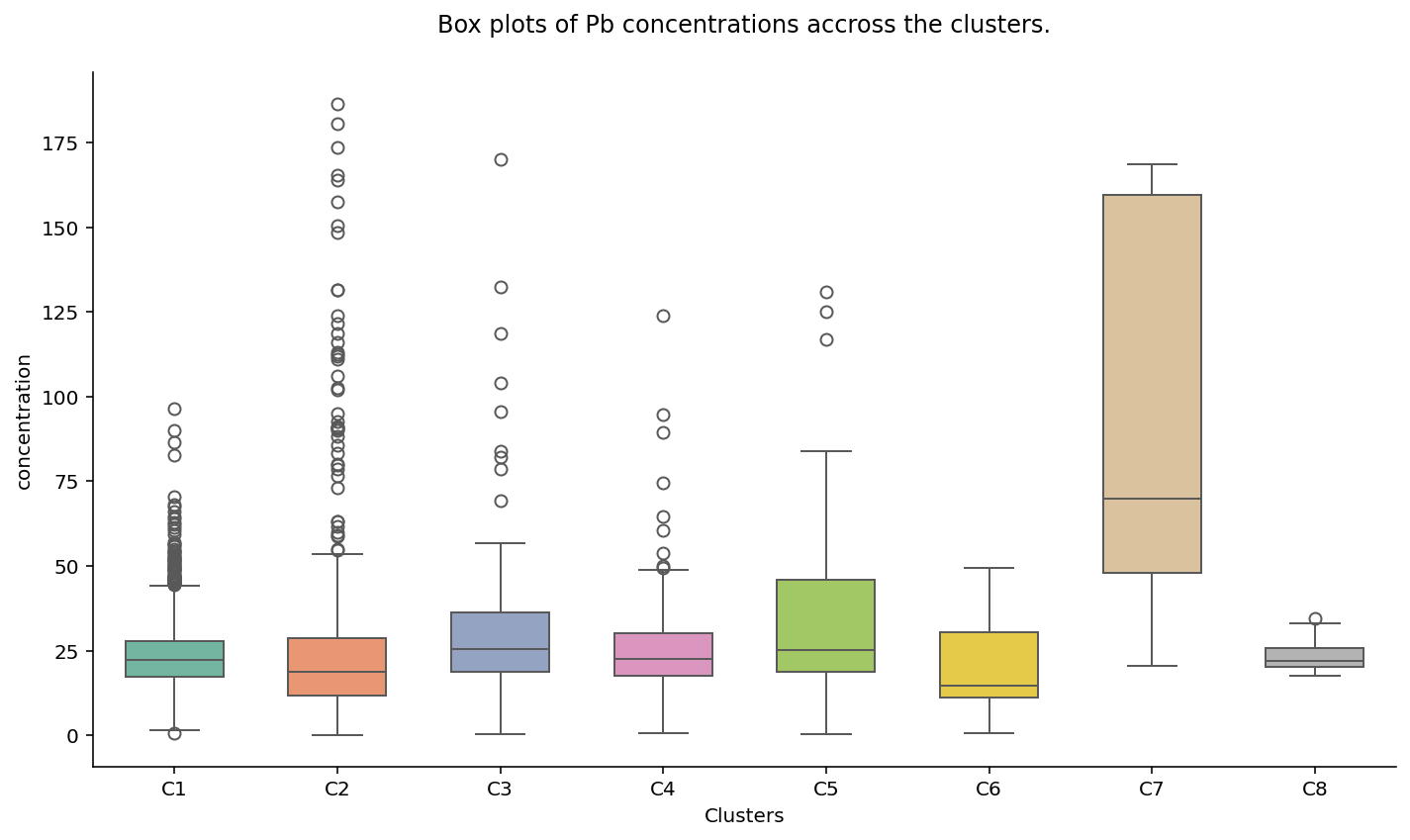}
    \includegraphics[width=0.475\textwidth]{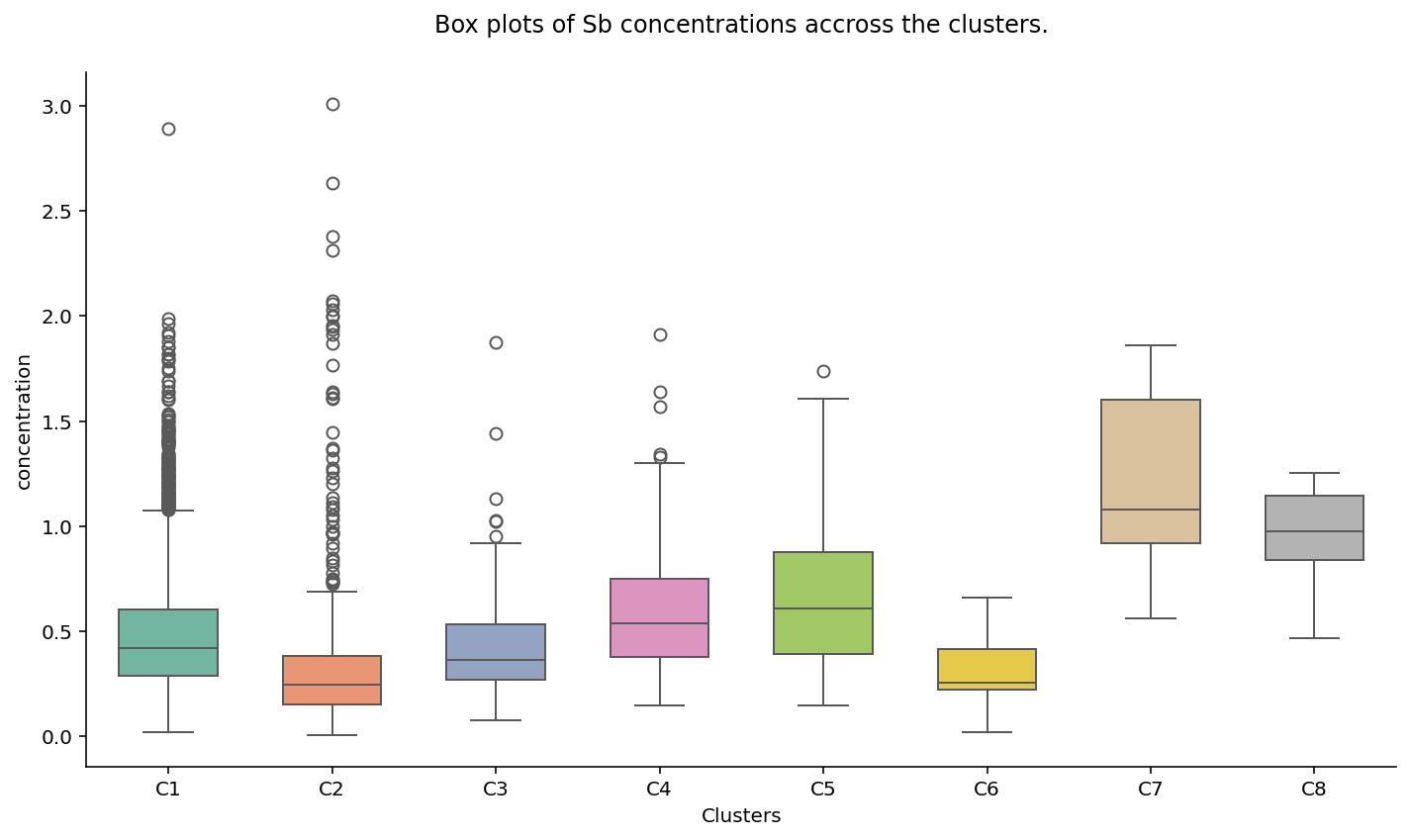}
    \includegraphics[width=0.475\textwidth]{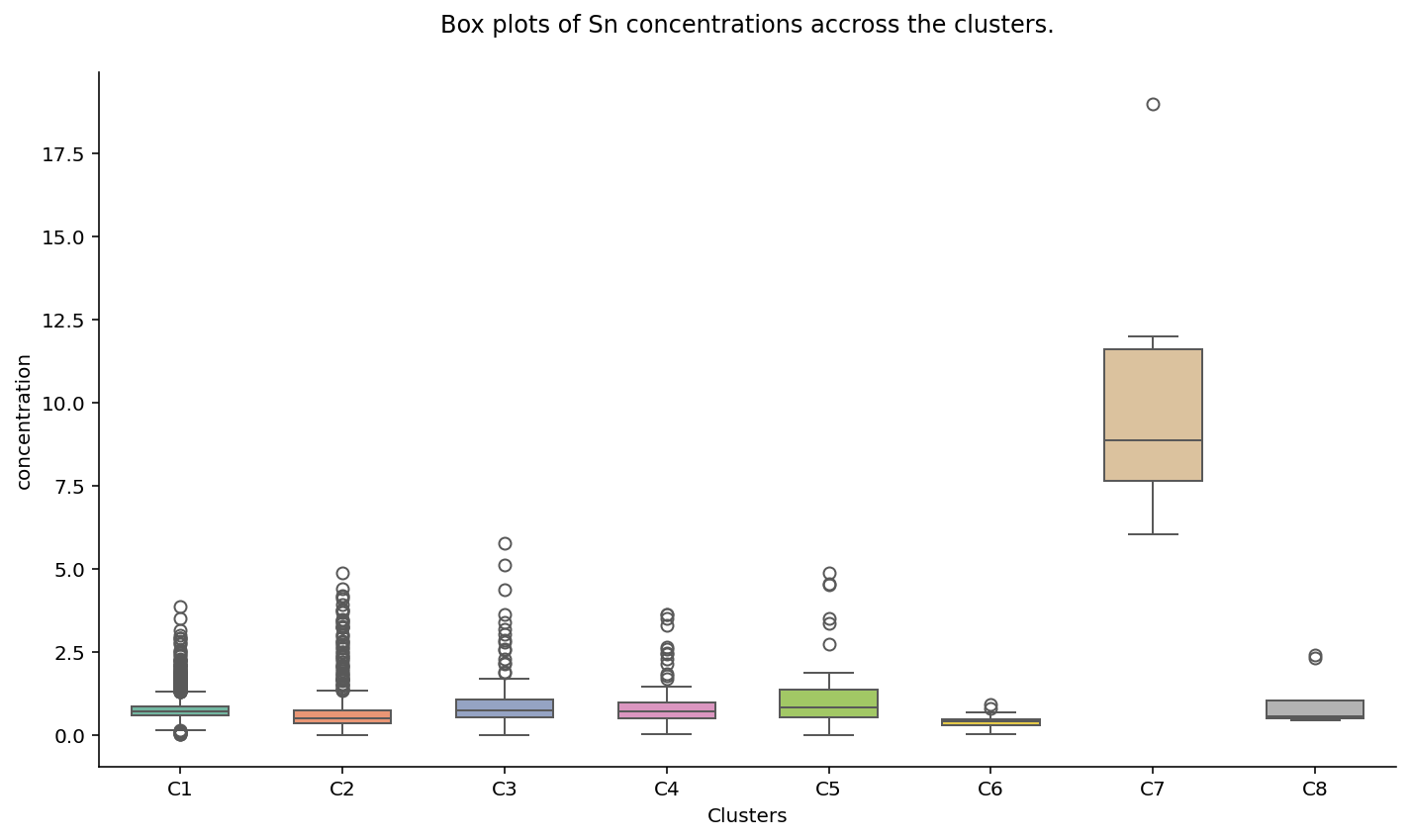}
    \includegraphics[width=0.475\textwidth]{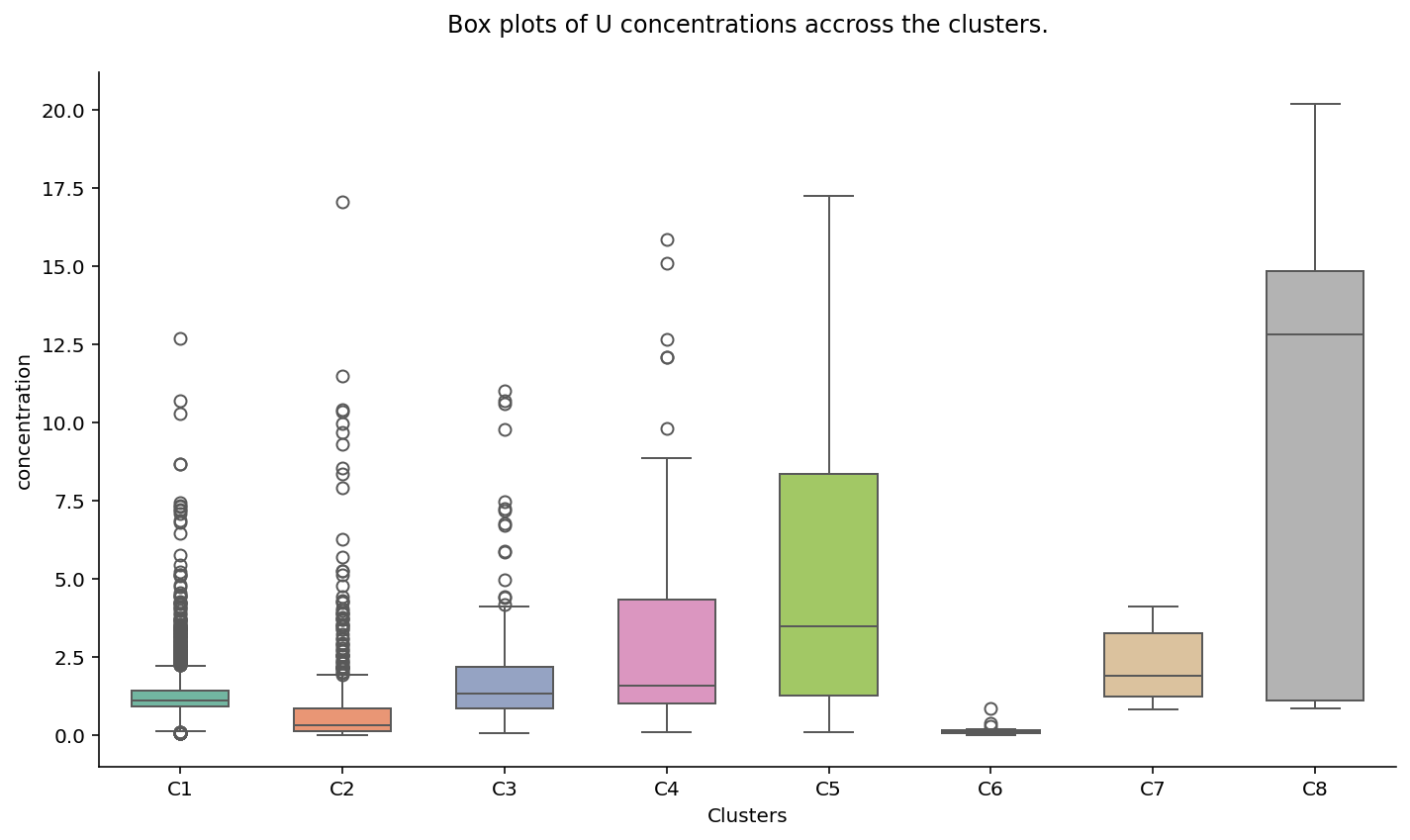}
    \includegraphics[width=0.475\textwidth]{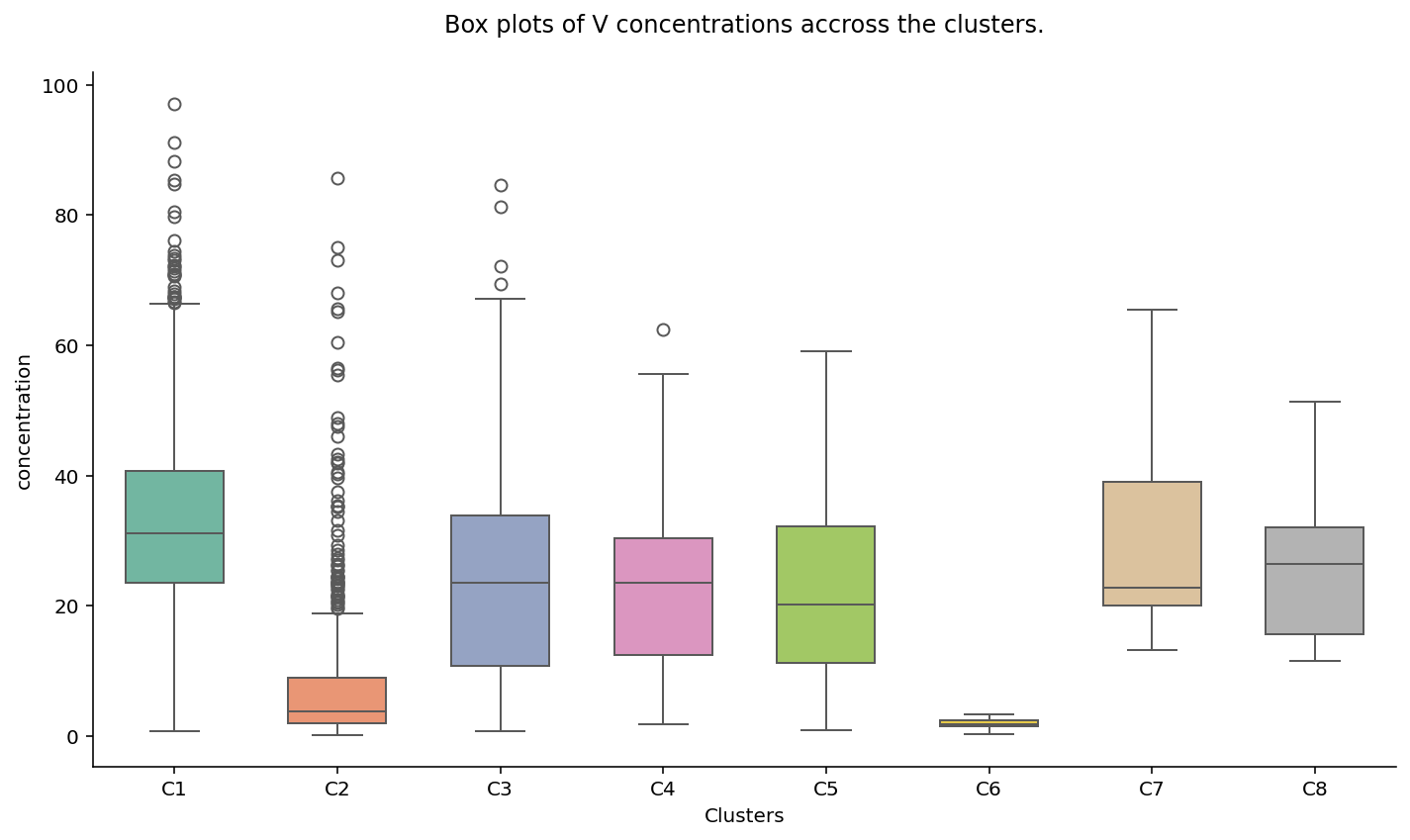}
    \includegraphics[width=0.475\textwidth]{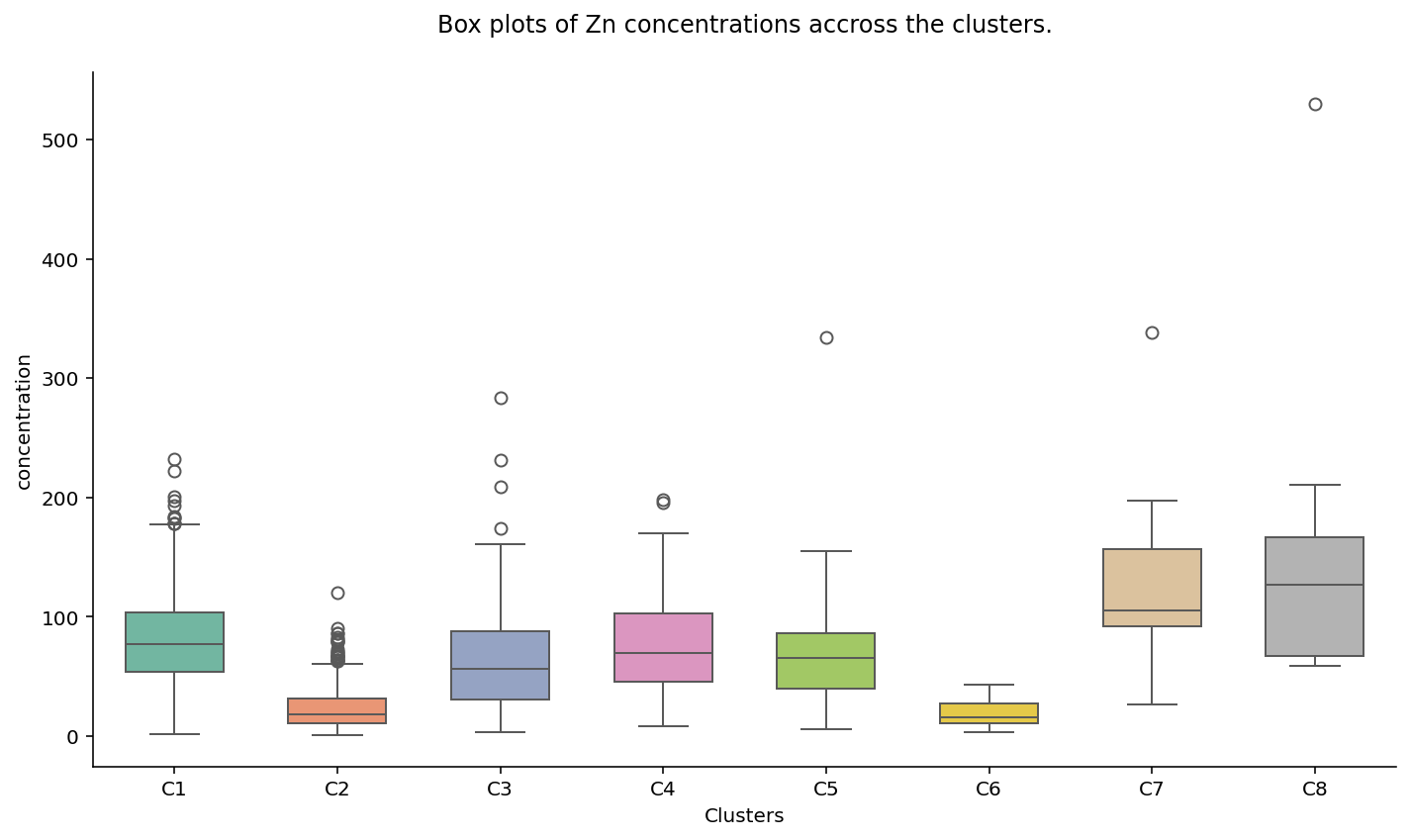}
	\caption{Box plots showing concentration distributions of the 15 PTEs across the eight identified soil clusters.}
\label{boxplot_all2}
\end{figure}

Figure \ref{boxplot_all1} and \ref{boxplot_all2} compare the box plots of each element among the eight soil clusters.
Key observations include:
\begin{itemize}
    \item Cluster 2 and Cluster 6 are characterized by the elements Mn, Ba, Zn, and Pb, which exhibit distinct concentration patterns compared to other clusters. While Mn, Ba, and Zn show high concentrations in all other clusters, their values are relatively low in Cluster 2 and Cluster 6. In contrast, Pb maintains a consistent concentration across all clusters, including the outlier set. Furthermore, the concentrations of the four elements (Mn, Ba, Zn, and Pb) show significantly higher values than the remaining 11 elements in Cluster 2 and Cluster 6. Figure \ref{C2&6_A} shows the geographic distribution of sampling sites for Cluster 2 and Cluster 6 soil samples. Figure \ref{C2&6_A} shows that the soil samples with low concentrations are primarily clustered in three regions. As revealed by the land cover map, soil samples in Cluster 2 and Cluster 6 are from wetlands and agricultural areas. The crescent-shaped feature of the central region requires further investigation.
\begin{figure}[!h]
	\centering
	\includegraphics[width=0.8\textwidth]{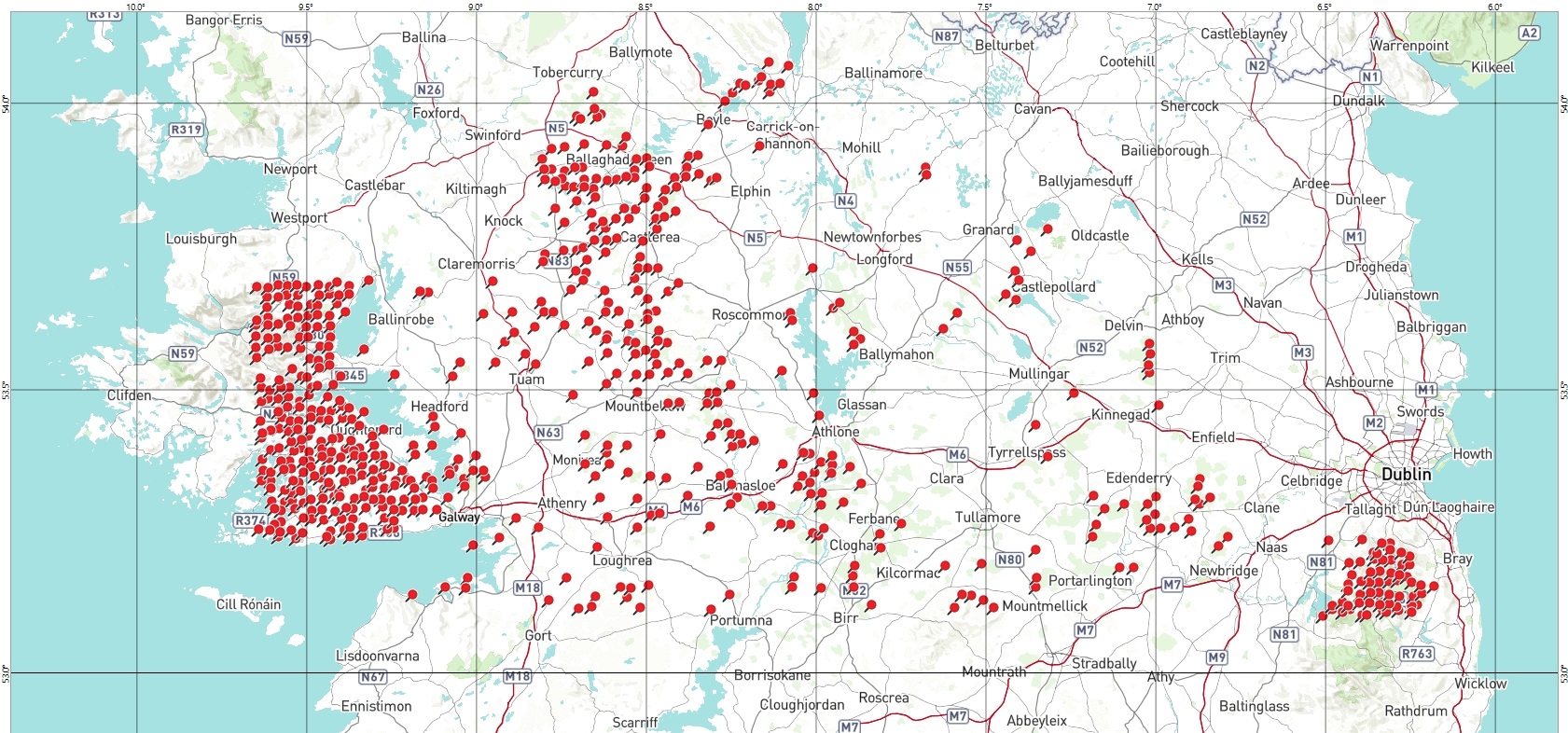}
         \includegraphics[width=0.8\textwidth]{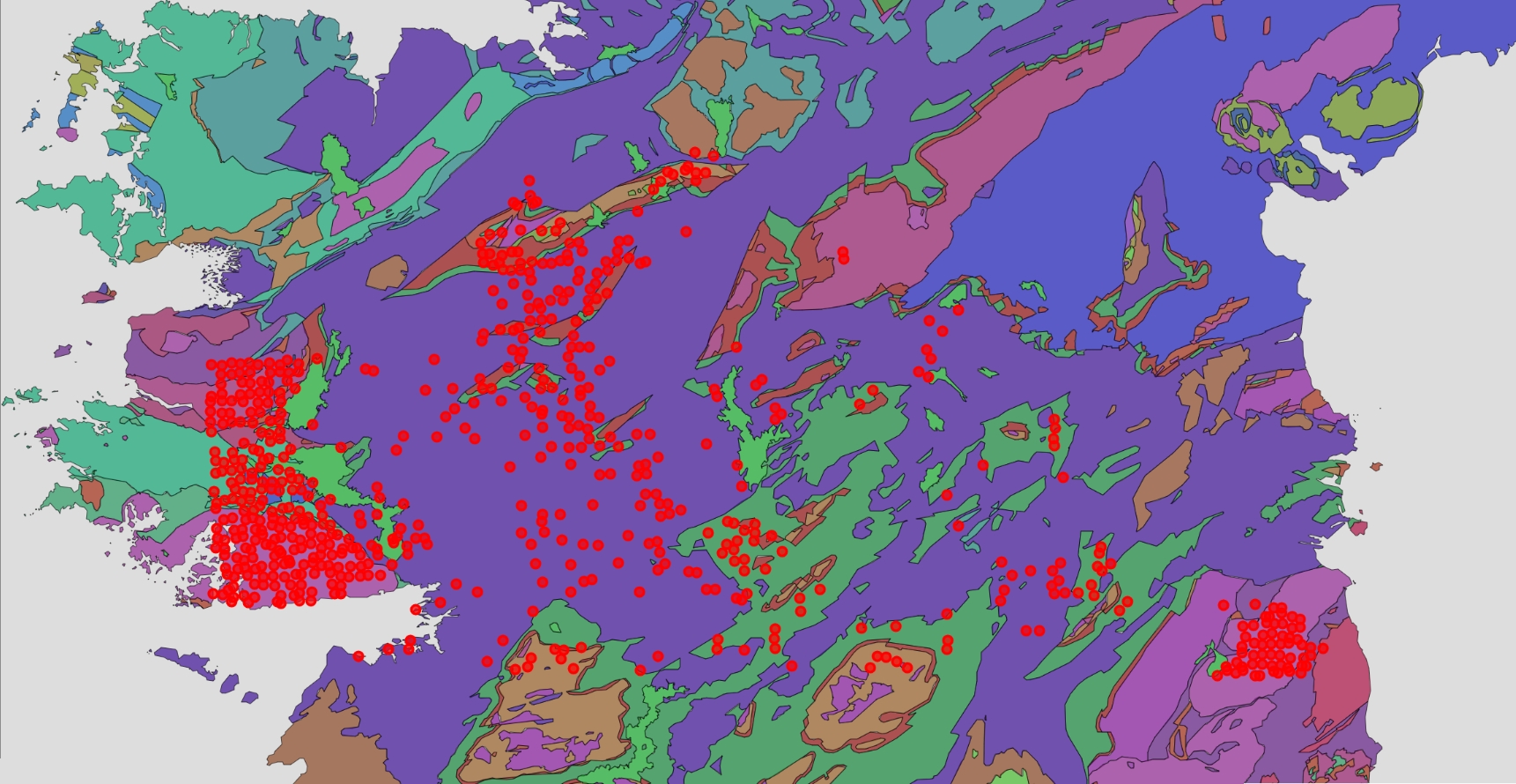}
         \includegraphics[width=0.8\textwidth]{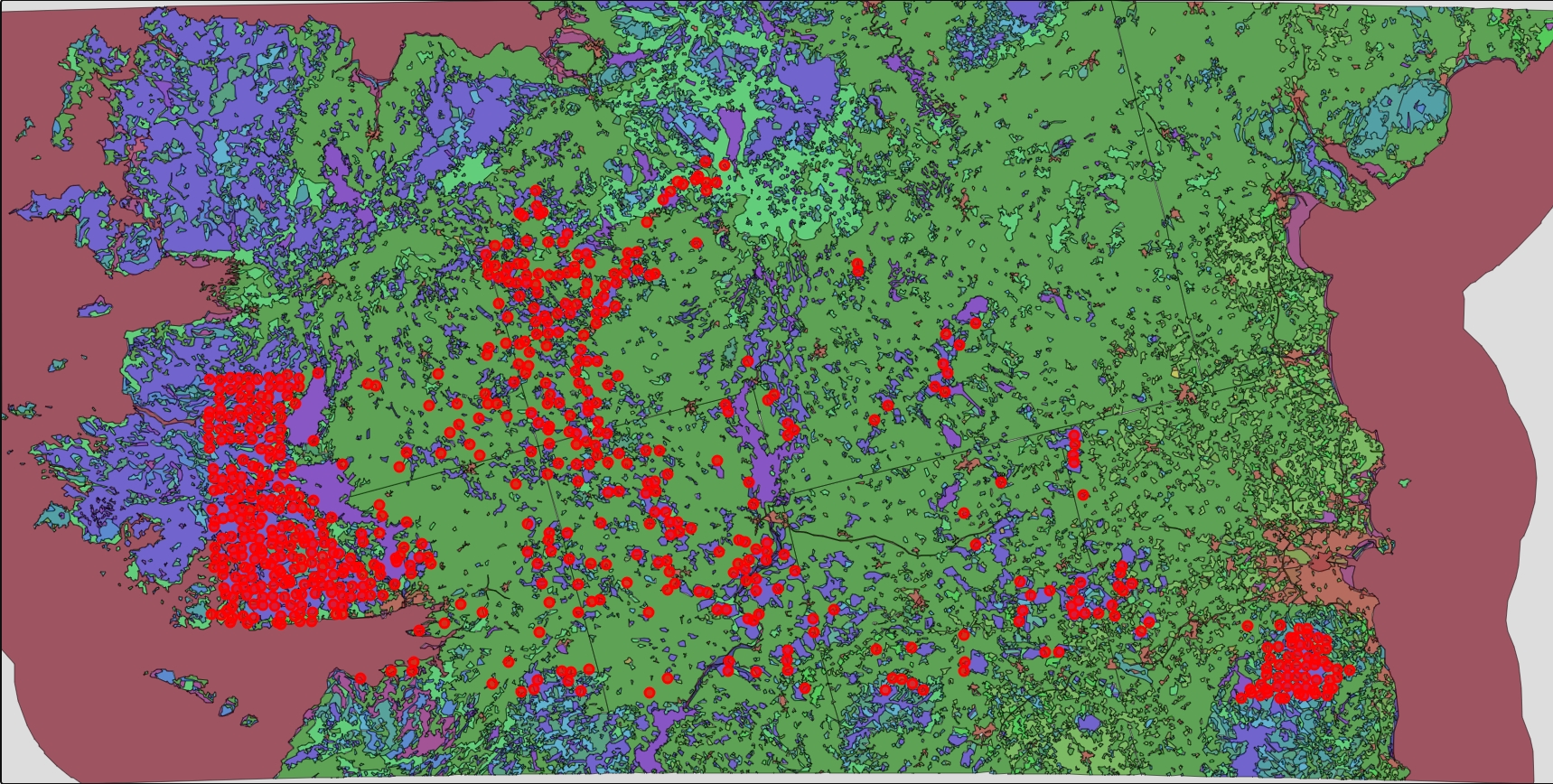}
	\caption{Geographic distribution of sampling sites for Cluster 2 and Cluster 6 soil samples. The background bedrock types are mainly limestone, granite, and sandstone. The background land cover types are mainly wetlands and agricultural areas.}
\label{C2&6_A}
\end{figure}
    \item Clusters 3, 4 and 5 exhibit similar elemental patterns, with Cluster 5 standing out through elevated median concentrations and greater variability in As, Ba, Mo, Sb, and U. Figure \ref{C345_T} shows the geographic distributions of soil samples across the three clusters. Cluster 3 is predominantly located in County Offaly, with a minor extension into Galway, while Cluster 4 is concentrated in County Westmeath. Cluster 5 displays a distinct crescent-shaped spatial feature.
\begin{figure}[!h]
	\centering
	\includegraphics[width=0.85\textwidth]{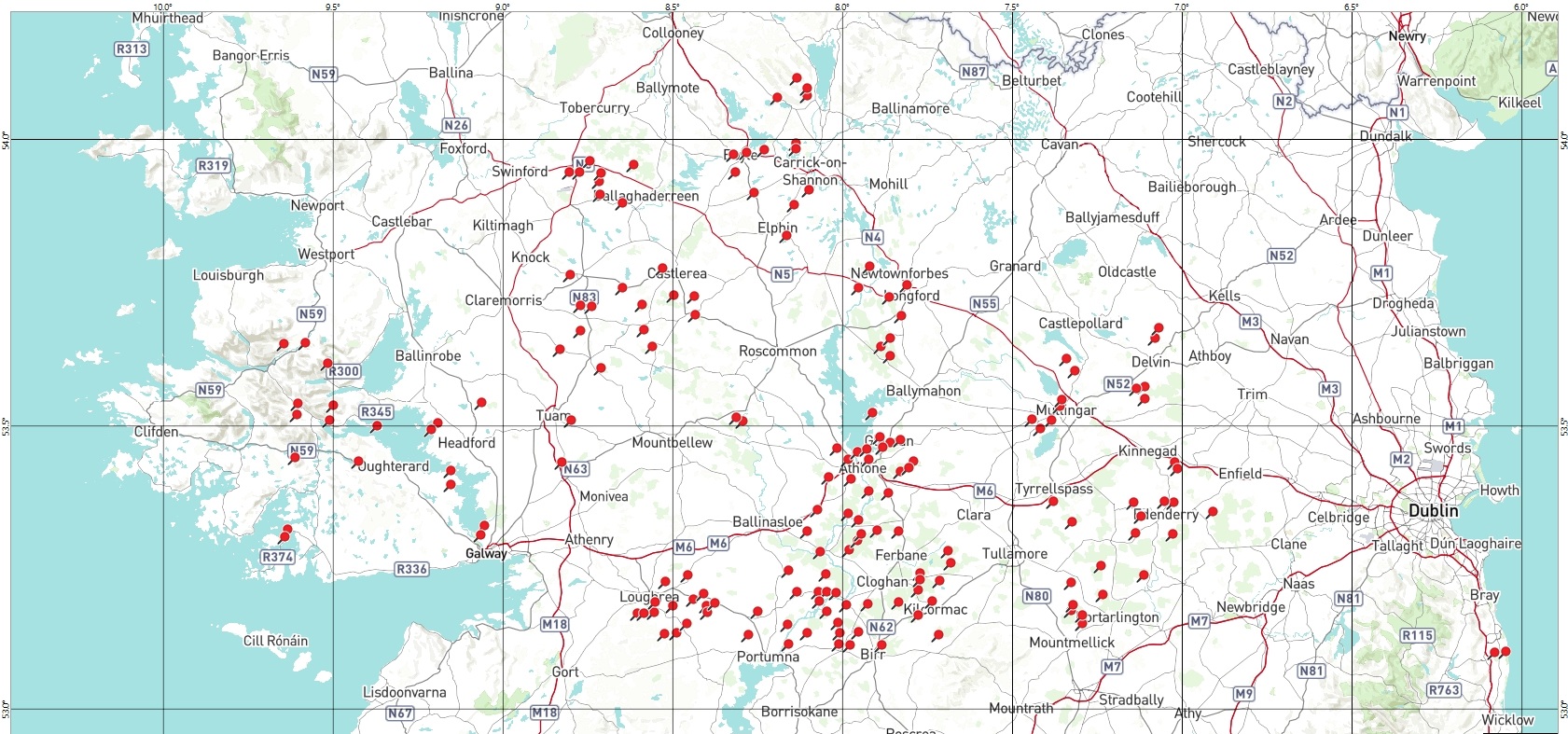}
         \includegraphics[width=0.85\textwidth]{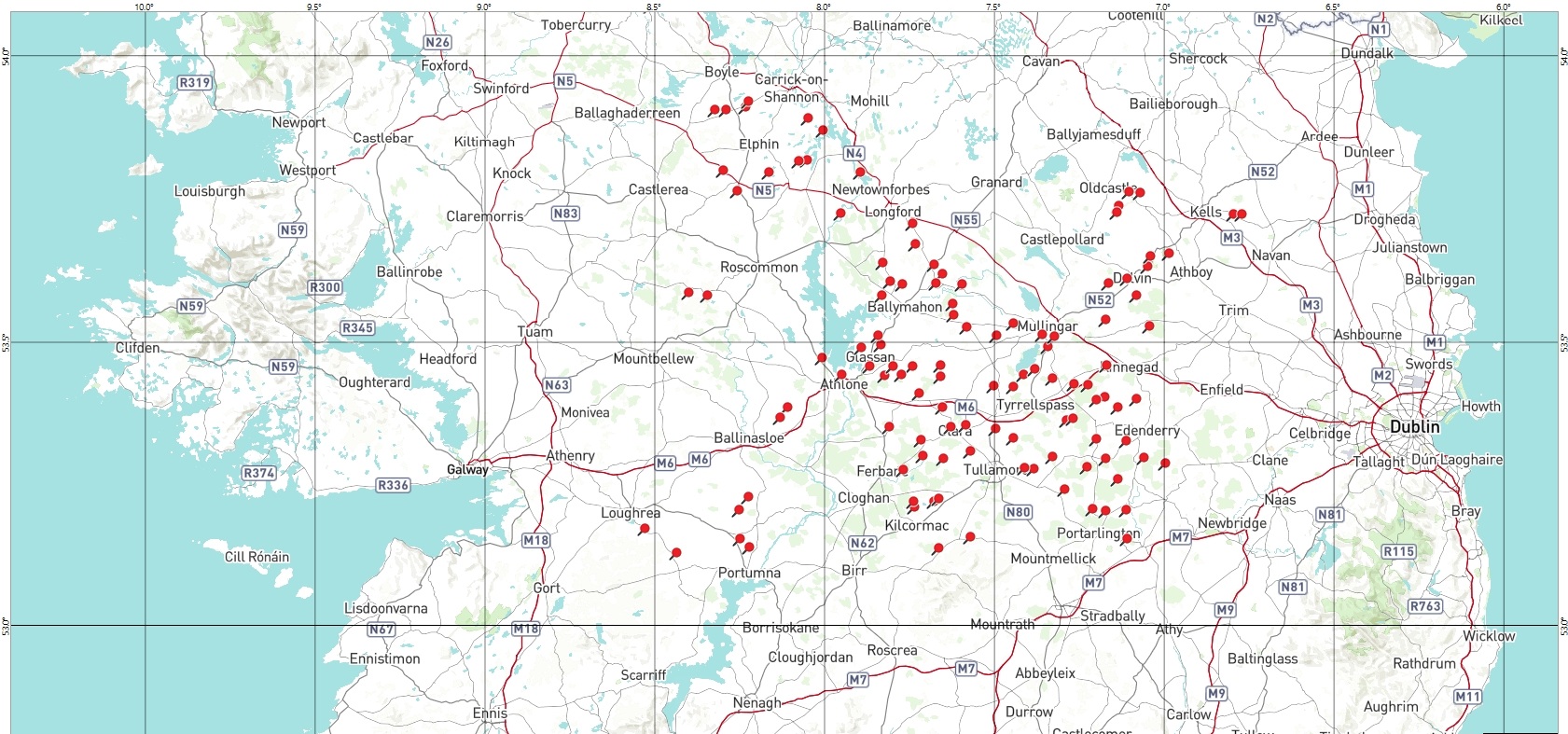}
         \includegraphics[width=0.85\textwidth]{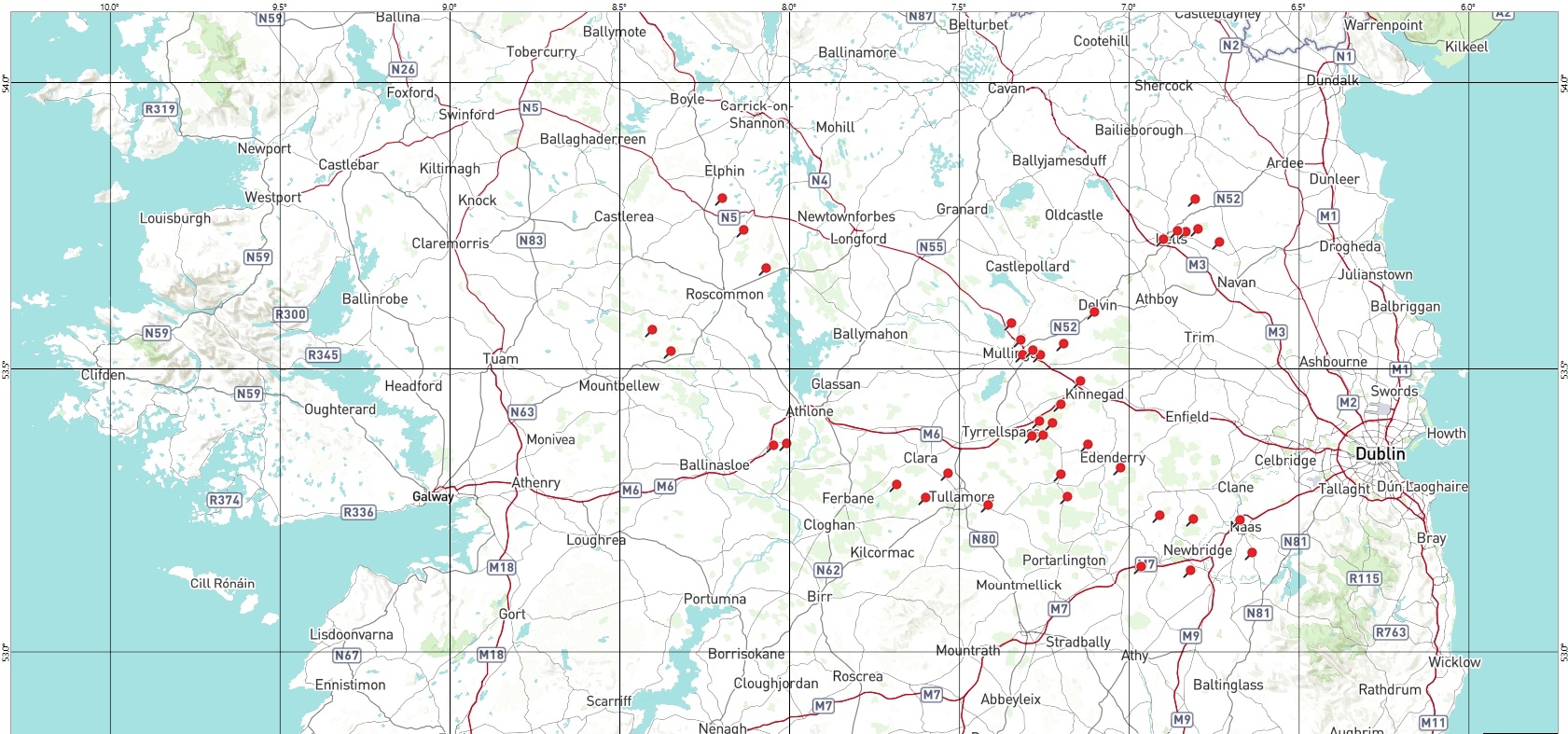}
	\caption{Geographic distribution of soil sampling sites for Cluster 3 (top), Cluster 4 (middle), and Cluster 5 (bottom).}
\label{C345_T}
\end{figure}
    \item Cluster 7 is characterized by Cu, Pb, and Sn, which exhibit significantly higher median concentrations and greater variability compared to their patterns in other clusters. Cluster 8 exhibits significant variability in As and U concentrations compared to other clusters. Soil samples in Cluster 7 \& 8 are visualized in Figure \ref{C78_AT}, where the second and third maps are respectively bedrock map and land cover map. Clusters 7 and 8 contain a total of 17 samples. The majority of these samples (15) are bedrock limestone, with the remaining two identified as deep marine greywacke, a type of sandstone. These soil samples are from areas characterized by wetlands and agricultural land cover.
\begin{figure}[htbp]
	\centering
	\includegraphics[width=0.85\textwidth]{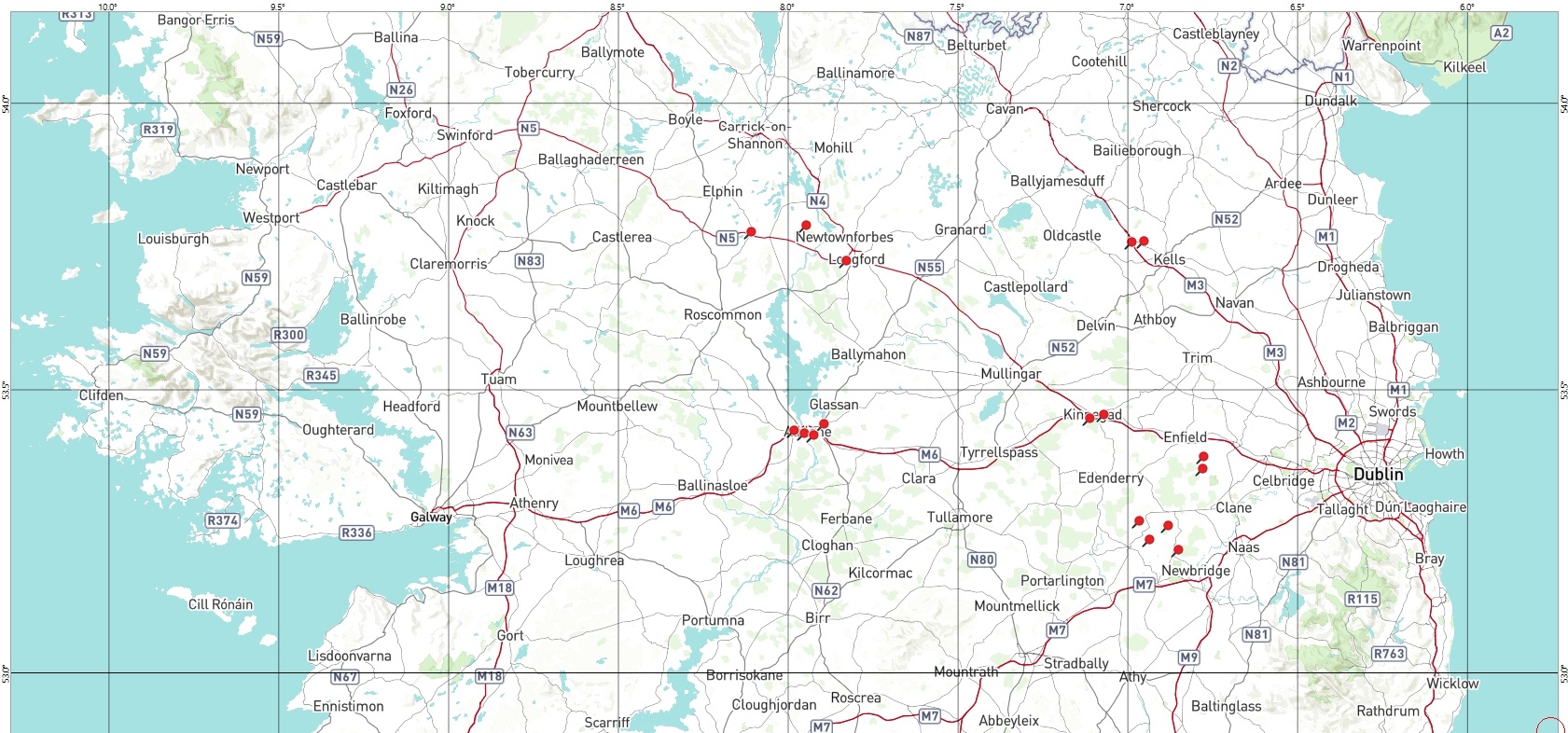}
         \includegraphics[width=0.85\textwidth]{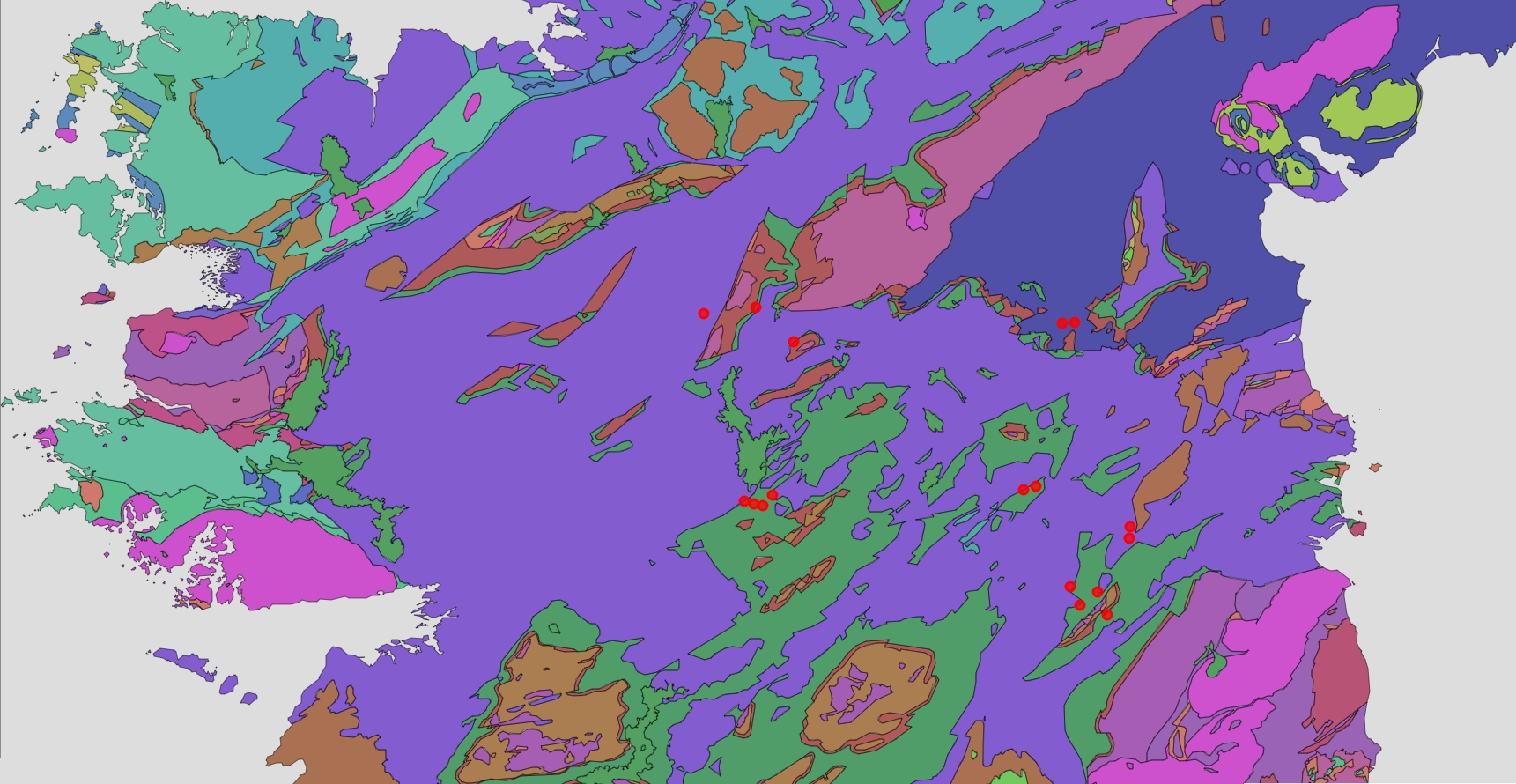}
         \includegraphics[width=0.85\textwidth]{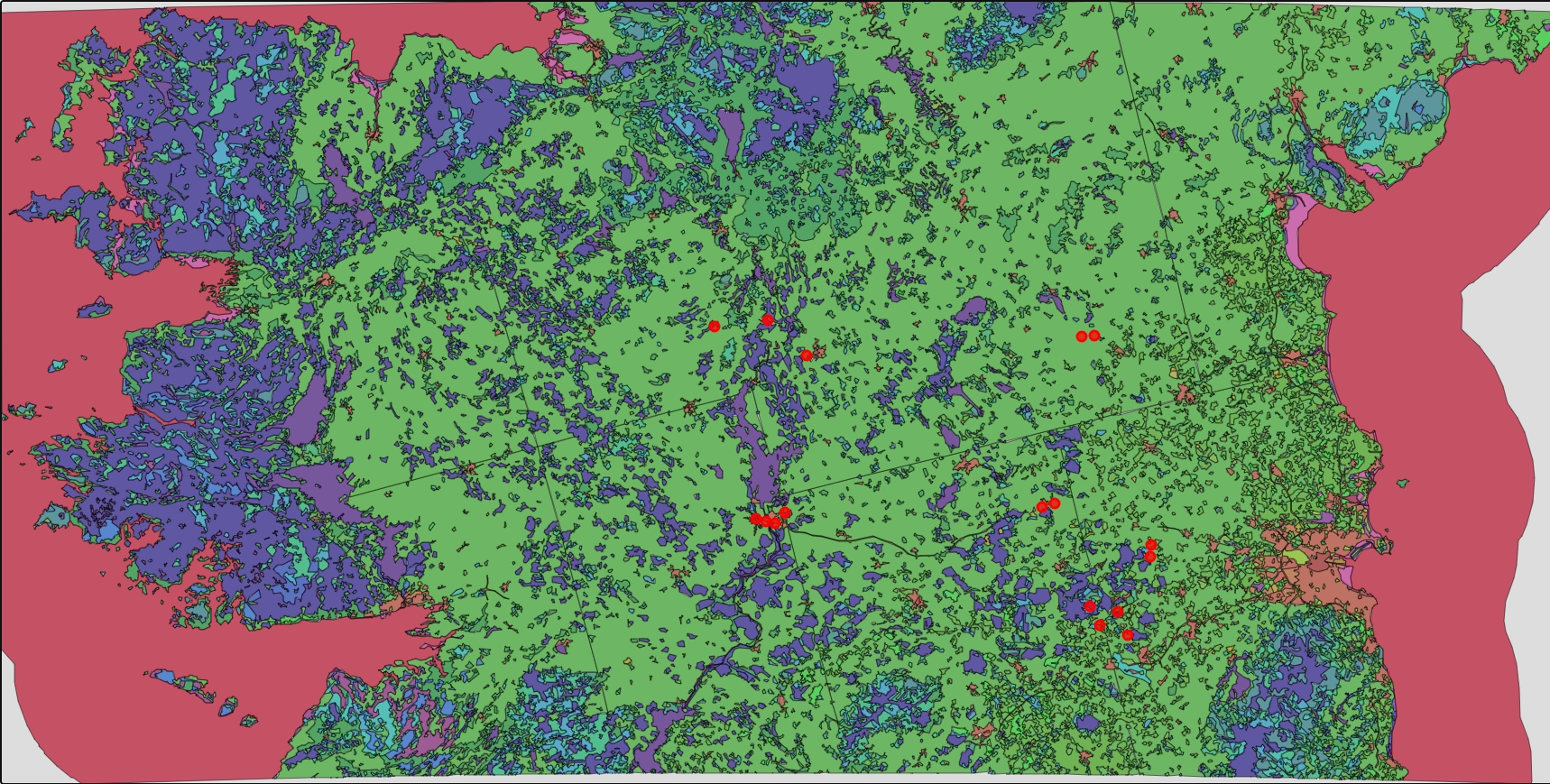}
	\caption{Geographic distribution of the sampling sites in Cluster 7 and Cluster 8.}
\label{C78_AT}
\end{figure}
\end{itemize}
The analysis divides the 15 PTEs into two distinct groups based on their spatial distribution patterns. The first group—comprising Ba, Cu, Mn, Zn, As, Mo, Pb, Sb, Sn, and U—shows significant variations in concentration across different areas of the investigated region. In contrast, the second group, which includes Cr, Ni, V, Bi, and Co, maintains consistent concentration levels throughout the entire study area.

\begin{figure}[htbp]
	\centering
	\includegraphics[width=0.8\textwidth]{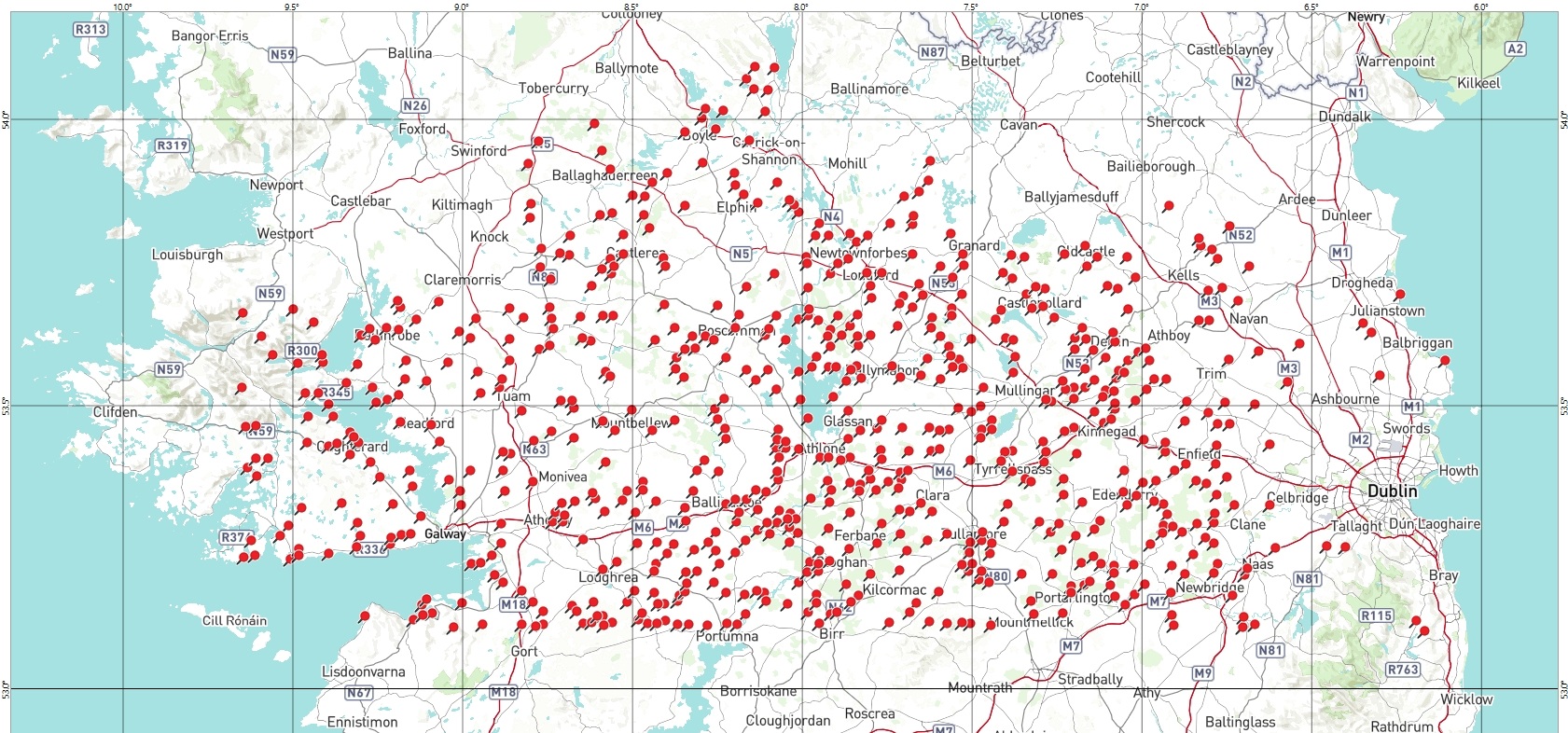}
         \includegraphics[width=0.8\textwidth]{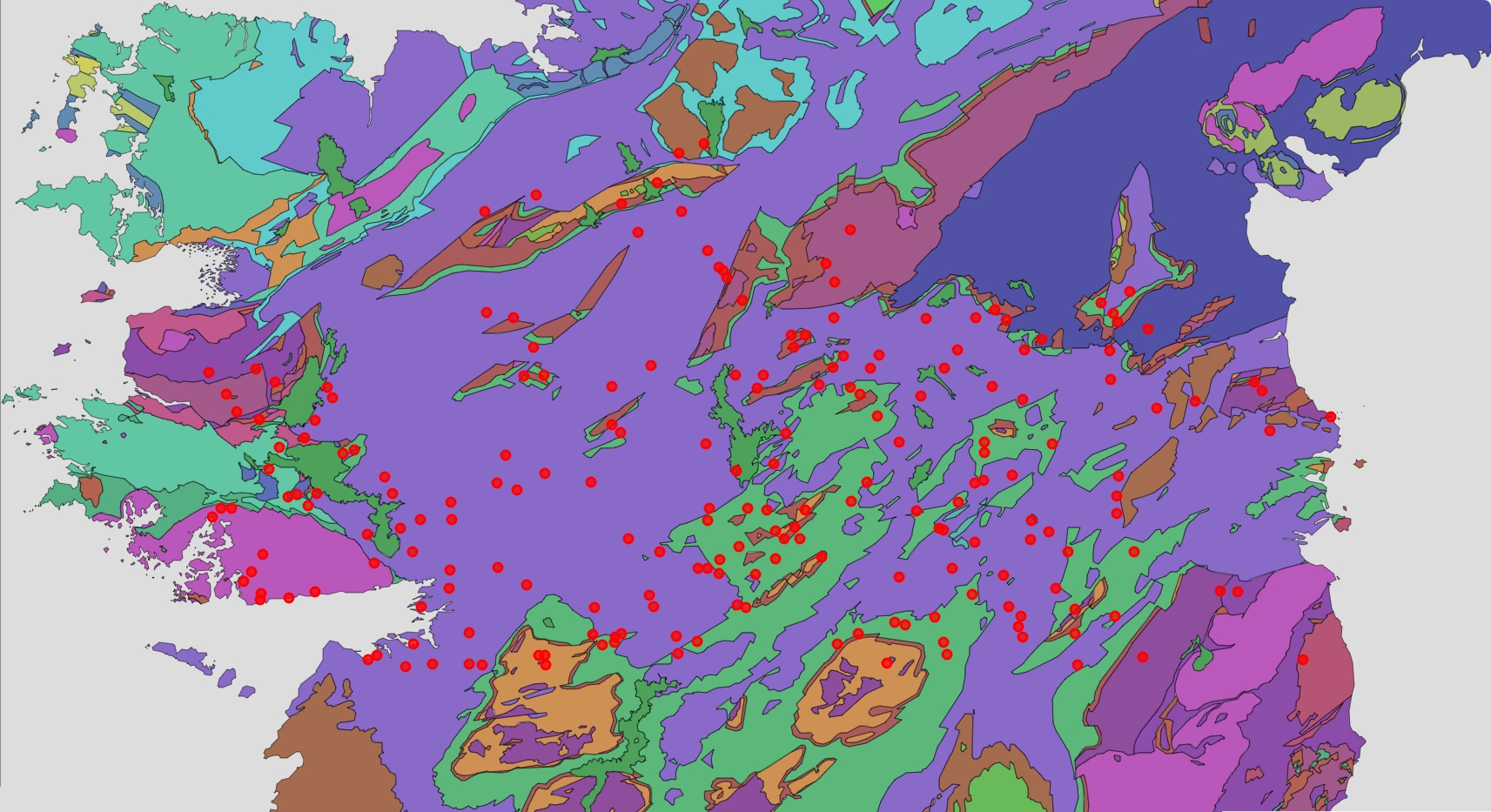}
         \includegraphics[width=0.8\textwidth]{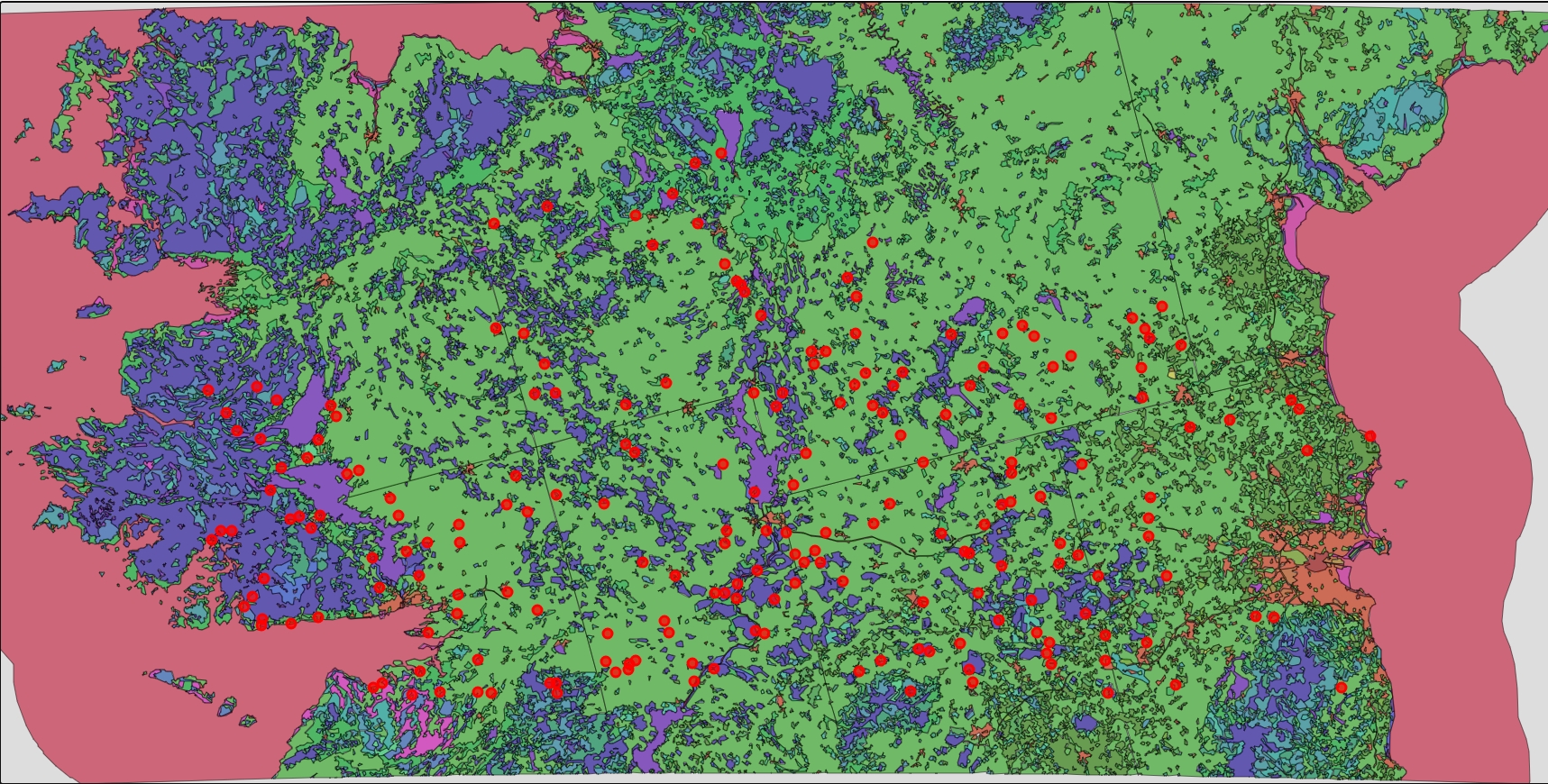}
	\caption{Geographic distribution of soil samples in the outlier set. Top: All outlier samples. Middle: 205 high-concentration outliers overlaid on bedrock geology. Bottom: The same 205 outliers mapped onto land cover types. }
\label{outlier_AT}
\end{figure}

Not all samples in the outlier set have extremely high concentrations; some are flagged as outliers because their multi-element composition differs significantly from neighboring samples. To identify abnormal samples based solely on concentration (independent of geographical factors), we applied the Isolation Forest method to the outlier set, setting the contamination parameter to 30\%. This method identified 205 outliers, all exhibiting exceptionally high concentrations of one or more PTEs. In Figure \ref{outlier_AT}, the top plot displays all soil samples in the outlier set, while the middle and bottom plots focus on the 205 outliers detected by Isolation Forest. The spatial distribution of these outliers closely mirrors that of the entire outlier set, with no nested patterns that might suggest large-scale contamination. However, the middle and bottom plots reveal that regions between longitude 9° to 9.5° (west of Ireland) and 7.5° to 8° show relatively higher contamination levels.

\section{Summary}
This study investigates topsoil contamination in Ireland using geochemical data from the Tellus Programme, analyzing 4,278 soil samples across 17,983 km². The research employs the spatial-CPF algorithm (the CPF clustering method with spatial constraints) to classify samples into eight clusters and an outlier set of 682 soil samples, revealing distinct contamination patterns.

Key findings include:
\begin{itemize}
    \item Cluster 1 represents healthy soils with low variability in potentially toxic elements.
    \item Clusters 2 and 6 also represent healthy soils. But they  exhibit unique Mn, Ba, Zn, and Pb patterns, differing from other clusters.
    \item Clusters 7 and 8 (17 samples) show elevated Cu, Pb, Sn, As, and U, linked to limestone bedrock and wetland/agricultural areas.
    \item Isolation Forest identified 205 extreme outliers with high PTE concentrations. Geographic hotspots were found between longitude 9°–9.5° (west of Ireland) and 7.5°–8°, though no large-scale contamination patterns were detected.
\end{itemize}

\section*{Acknowledgment}
I would like to express my gratitude to Áine Sadlier for her valuable preliminary exploratory work, which was conducted as part of her final-year project.

\bibliographystyle{apalike}
\bibliography{Ref_nLRL}
\end{document}